\documentstyle[prl,aps,eqsecnum,twocolumn,12pt]{revtex}

\newcommand{\id}{{1\!\!1}} 
\def\e{{\rm e}\,}
\newcommand{\rf}[1]{(\ref{#1})}
\newcommand{\eq}[1]{Eq.~(\ref{#1})}
\newcommand{\non}{\nonumber \\*}

\def\be{\begin{equation}}
\def\ee{\end{equation}}
\def\bea{\begin{eqnarray}}
\def\eea{\end{eqnarray}}

\def\d{\partial}

\def\v{u}
\def\tr{{\rm tr}}

\font\mybb=msbm10 at 12pt
\def\bb#1{\hbox{\mybb#1}}

\newcommand{\zed}{{\bb Z}} 
\newcommand{\real}{{\bb R}} 

\begin{document}
\draft

\baselineskip=14pt

\title{String Theory in Electromagnetic Fields}

\author{J. Ambj\o rn$^a$, Y.M. Makeenko$^b$, G.W. Semenoff $^c$ and R.J.
Szabo$^d$}

\address{\baselineskip=12pt \hbox{~} \hbox{$^a$ The Niels Bohr Institute,
Blegdamsvej 17, DK-2100 Copenhagen \O, Denmark} \hbox{\tt ambjorn@nbi.dk}}

\address{\baselineskip=12pt \hbox{$^b$ Institute of Theoretical and
Experimental Physics, B. Cheremushkinskaya 25, 117218 Moscow, Russia}
\hbox{\tt makeenko@itep.ru}}

\address{\baselineskip=12pt \hbox{$^c$ Department of Physics and Astronomy,
University of British Columbia, 6224 Agricultural Road,} \hbox{Vancouver, B.C.
V6T 1Z1, Canada} \hbox{\tt semenoff@physics.ubc.ca}}

\address{\baselineskip=12pt \hbox{$^d$ Department of Mathematics, Heriot-Watt
University, Riccarton, Edinburgh EH14 4AS, Scotland} \hbox{\tt
R.J.Szabo@ma.hw.ac.uk} \hbox{~}}

\date{\baselineskip=12pt \hbox{Preprint ITEP--TH--78/00 ,
HWM00--27 , hep-th/0012092 , December 2000}}

\maketitle

\section*{ABSTRACT}

\begin{abstract}
\baselineskip=12pt
\noindent
A review of various aspects of superstrings in background electromagnetic
fields is presented. Topics covered include the Born-Infeld action, spectrum of
open strings in background gauge fields, the Schwinger mechanism,
finite-temperature formalism and Hagedorn behaviour in external fields, Debye
screening, D-brane scattering, thermodynamics of D-branes, and noncommutative
field and string theories on D-branes. The electric field instabilities are
emphasized throughout and contrasted with the case of magnetic fields. A new
derivation of the velocity-dependent potential between moving D-branes is
presented, as is a new result for the velocity corrections to the one-loop
thermal effective potential.
\end{abstract}

\setcounter{section}{0}

\section{Introduction}

Superstring theory is a fundamental candidate for a theory of quantum gravity
because its elementary closed string spectrum naturally induces background
fields of ten-dimensional supergravity. Among the bosonic fields one finds, in
addition to the metric tensor of ten-dimensional spacetime, a torsion
Neveu-Schwarz two-form field as well as higher and lower degree differential
form
Ramond-Ramond fields. The former field, when it is closed or equivalently
on-shell, is formally equivalent to a background
electromagnetic field strength tensor in spacetime, while the latter ones are
the objects which couple to D-branes, the extended hyperplanes in spacetime
onto which open strings attach (with Dirichlet boundary conditions).
D$p$-branes are $p$-dimensional soliton-like objects whose quantum dynamics are
described by the quantum theory of the open strings whose ends are constrained
to move on them. In the low-energy limit, there are a finite number of massless
fields which survive whose dynamics are described by a $p+1$-dimensional
effective quantum field theory. One of these fields is a $U(1)$ gauge field.
Therefore, understanding the behaviour of strings and D-branes in the presence
of electromagnetic fields is important for the description of non-perturbative
vacuum states in superstring theory. Furthermore, using duality, this problem
is also important for understanding various aspects of D-brane dynamics.

In this review article we shall focus on a particularly tractible problem, that
of open strings and D-branes in constant background electric and magnetic
fields. These models have attracted renewed interest very recently because they
give an explicit realization of some old conjectures about the nature of
spacetime at very short distance scales. If one is to use string states as
probes of short distance structure, then one cannot probe lengths smaller than
the intrinsic length of the strings. Therefore, below the string scale the
notion of geometry must drastically change, and an old proposal is that the
spacetime coordinates become noncommuting operators. The deformation of D-brane
worldvolumes to noncommutative manifolds by the external electromagnetic field
has led to a revival of interest in these earlier suggestions. In addition, the
effective low-energy dynamics can be described by new, noncommutative versions
of ordinary quantum field and string theories, and hence a wealth of new
problems for both field theorists and string theorists.

Motivated by these issues, in the following we will present an overview of some
of the fundamental aspects of string theory in electromagnetic fields. The
qualitative effects can all be seen at the level of the simpler bosonic string
theory, which we will confine most of our attention to in this paper. As we
indicate throughout, the results readily extend to the case of superstrings.
Many of the novel effects exhibited by strings in background fields can be seen
at the level of free open strings, or equivalently (in Type IIB superstring
theory) for D9-branes which fill the spacetime. This is the topic of section
II. We will derive the effective gauge field dynamics for the open strings up
to one-loop order in string perturbation theory, and describe the spectrum of
the string theory. We shall also start seeing here some important differences
between electric and magnetic backgrounds in superstring theory. While strings
in external magnetic fields possess no more instabilities than the quanta of
Yang-Mills gauge theory, electric backgrounds play a much different role in
string theory. In addition to the usual instability of the vacuum in an
electric background that occurs in quantum electrodynamics, strong electric
fields can tear apart a string and render both the classical and quantum
theories physically meaningless.

As we have mentioned, string theory exhibits a variety of novel effects at very
high energies. Non-trivial background fields may also have an effect on the
properties of strings in this regime. In particular, one can examine how the
external fields modify the behaviour of strings at high temperatures, where
they are known to undergo a phase transition into a sort of deconfining phase
in which the strings propagate as long string states in the spacetime. Free
strings at finite temperature and in background electromagnetic fields will be
analysed in section III.

One of the most important applications of the external field problem for free
open strings is its interpretation in the $T$-dual picture, where it maps onto
the problem of moving D-branes. This problem is dealt with at length in section
IV. Here we present a new derivation, which contains some novel technical
details that may be of use for other calculations, of the well-known scattering
amplitude between two D-branes travelling at constant velocity. The
corresponding thermodynamic problem is particularly interesting in this case. A
special class of black holes in string theory admits a dual description as a
configuration of D-branes. By using the quantum string theory living on the
D-brane, one can compute the Bekenstein-Hawking entropy and the rate of thermal
radiation from the black hole. The corresponding Hawking temperature is
conjectured to be the same in this case as the extrinsic temperature of a
Boltzmann gas of D-branes. These features of the thermal ensemble of D-branes
can be checked by computing the free energy using the effective, low energy
description of D-brane dynamics in terms of supersymmetric Yang-Mills theory
with 16 supercharges. In section IV we shall also present new results for the
leading velocity corrections to the one-loop thermal potential between
D-branes.

The final instance of the constant external field problem, which we address in
section V, is to study the properties of D-branes themselves in the
electromagnetic background. Here we shall focus on the geometric modifications
that are caused by the external field. We shall see that, generically, the
D-brane worldvolume is not a conventional manifold and is described by a
noncommutative space. This is again a particular effect of the quantum open
string theory that lives on the D-branes. Here we shall see a particularly
drastic distinction between electric and magnetic fields. In a particular
low-energy limit, the effective dynamics of the noncommutative D-branes is
described in the magnetic case by a deformation of the usual gauge field
dynamics on the branes, while in the electric case there is no field theory
limit and the effective theory is a deformation of the usual open string theory
on the D-branes. In this latter case, the noncommutativity is given directly in
terms of the string scale, and the most intesting aspect of this open string
theory on the noncommutative manifold is that it does not contain closed
strings. In particular, it is a novel example of a string theory which does not
contain gravity. We can expect that these theories capture many of the
important features of the standard string theories, but without being plagued
by the conceptual problems that arise due to the presence of gravitation.

\section{Open strings in background gauge fields}

In this section we will start describing some of the basic physical properties
of strings in an external electromagnetic field. An external gauge field
couples to an open string through Chan-Paton factors at
the string endpoints. Therefore, because of the Green-Schwarz anomaly
cancellation condition, all of our considerations in this section and
the next strictly speaking only apply to Type I superstrings, since Type II
superstring theory has no gauge group. The gauge field is then associated with
a subgroup of the $SO(N)$ gauge group of Type I string theory, where
$N=2^{d/2}$ and $d$ is the dimension of spacetime which we will assume is even.
By an electromagnetic field we will mean one that is associated with an abelian
subgroup of this gauge group. However, we will only write down explicit
formulas which also pertain in principle to Type II superstrings, as they will
become relevant in sections IV and V when the open strings will attach to
D-branes which can host electromagnetic fields in the guise of a Neveu-Schwarz
two-form field.

\subsection{The Born-Infeld action}

In this subsection we will derive the low-energy effective action which governs
the propagation of free open strings in a slowly-varying background
electromagnetic field $F_{\mu\nu}$ \cite{FT,ACNY,CLNY1} (See \cite{Tseytrev}
for a review). In string perturbation theory and in the RNS formulation, the
vacuum energy may be computed in first quantization from the Polyakov path
integral
\bea
{\cal Z}[F]&=&-\sum_{h=0}^\infty g_s^{2h-1}\,\sum_\sigma\,\int Dg_{ab}~
D\chi_{ab}\non&&\times\,\int Dx^\mu~D\psi^\mu~\e^{-S[g,x,\chi,\psi;A]} \ ,
\label{calZF}\eea
where $g_s$ is the string coupling constant whose powers weight the genus $h$
of the open string worldsheet which has Euclidean metric $g_{ab}$ (and
superpartner the two-dimensional gravitino field $\chi_{ab}$), and the sum
over spin structures $\sigma$ with the appropriate weights imposes the GSO
projection that leads to modular invariance, a tachyon-free spectrum, and
spacetime supersymmetry of the string theory. We will assume in this section
that the target space has flat Euclidean metric $\delta_{\mu\nu}$.

The first contribution to the partition function (\ref{calZF}) comes from the
disc diagram $\Sigma$, which by conformal invariance of the classical theory
can be parametrized by coordinates $z=r\,\e^{i\vartheta}$ with $0\leq r\leq1$
and
$0\leq\vartheta\leq2\pi$. The bosonic part of the tree level string action in
the
conformal gauge is
\bea
S_0[x,A]&=&\frac1{4\pi\alpha'}\,\int d^2z~\partial x^\mu\,\overline{\partial}
x_\mu\non&&
-\,ie\left.\int\limits_0^{2\pi}d\vartheta~\dot x^\mu\,A_\mu(x)\right|^{r=1} \ ,
\label{Sdisc}\eea
where here and in the following a dot will denote differentiation with respect
to the worldsheet boundary coordinate $\vartheta$. The quantity
$T_s=1/2\pi\alpha'$ is the string tension. The endpoints of the open strings
carry charges $e$ which couple to the electromagnetic vector potential
$A_\mu(x)$. When inserted into (\ref{calZF}), the action (\ref{Sdisc}) leads to
the expectation value, with respect to the free worldsheet $\sigma$-model (the
bulk term in (\ref{Sdisc})), of the Wilson loop operator for the gauge field
$A_\mu$ over the boundary of $\Sigma$. To evaluate it, we use the background
field approach and expand the string embedding fields as
$x^\mu=x_0^\mu+\xi^\mu$, where $x_0^\mu$ are their constant zero modes. The
tree-level contribution to (\ref{calZF}) then involves the propagator
$G^{\mu\nu}(z,z')=\langle0|T[\xi^\mu(z)\xi^\nu(z')]|0\rangle
=-2\pi\alpha'\,\delta^{\mu\nu}\,N(z,z')$, where
\be
N(z,z')=\frac1{2\pi}\,\ln\Bigl|z-z'\Bigr|\left|z-\overline{z}'^{-1}\right|
\label{Neumanndisc}\ee
is the Neumann function for the disc which satisfies the equation of motion
$\nabla^2N(z,z')=\delta(z-z')$ and the Neumann boundary condition
$\partial_rN(z,z')|_{r=1}=0$.\footnote{\baselineskip=12pt This Green's function
may be derived by using the method of images after mapping the disc into the
upper complex half-plane via a conformal transformation.} On the boundary of
the worldsheet, where
$z=\e^{i\vartheta}$, the Green's function (\ref{Neumanndisc}) becomes
\bea
N(\vartheta,\vartheta')&=&\frac1{2\pi}\,\ln
\Bigl(2-2\cos(\vartheta-\vartheta')\Bigr)
\non&=&-\frac1\pi\,\sum_{n=1}^\infty\frac{\cos n(\vartheta-\vartheta')}n~
\e^{-\varepsilon n} \ ,
\label{Nboundary}\eea
where $\varepsilon\to0^+$ is an ultraviolet cutoff which regulates the
logarithmic short-distance $\vartheta\to\vartheta'$ singularity of
(\ref{Nboundary}),
$N(\vartheta,\vartheta)=\frac1\pi\,\ln\varepsilon$. We will work in the radial
gauge
$\xi^\mu A_\mu(x_0+\xi)=0$, $A_\mu(x_0)=0$, and with slowly varying gauge
fields which admit an expansion
$A_\mu(x_0+\xi)=\frac12\,F_{\mu\nu}(x_0)\,\xi^\nu+{\cal O}(\partial F)$. In the
following we will evaluate the vacuum amplitude to leading orders in the
expansion in derivatives of the field strength tensor $F_{\mu\nu}$.

After integrating out the bulk values of the string coordinates in the interior
of the disc, the bosonic sector of the Polyakov path integral (\ref{calZF}) at
tree-level and in the conformal gauge becomes
\be
Z_0[F]=\frac1{g_s}\,\int d\vec x_0~\int D\xi^\mu~\e^{-S_\partial[\xi,A]} \ ,
\label{ZFpathint}\ee
where the effective boundary action is
\bea
S_\partial[\xi,A]&=&\frac12\,\int\limits_0^{2\pi}d\vartheta~\left(\frac1
{2\pi\alpha'}\,\xi^\mu\,N^{-1}\,\xi_\mu\right.\non&&+\biggl.
ieF_{\mu\nu}\,\xi^\mu\,\dot\xi^\nu\biggr) \ .
\label{Sboundary}\eea
Here $N^{-1}$ denotes the coordinate space inverse of the boundary Neumann
function (\ref{Nboundary}) which is given explicitly by
\be
N^{-1}(\vartheta,\vartheta')=-\frac1\pi\,\sum_{n=1}^\infty n
\cos n(\vartheta-\vartheta') \ ,
\label{Ninverse}\ee
where we have used the completeness relation
\be
\frac1\pi\,\sum_{n=1}^\infty\cos
n(\vartheta-\vartheta')=\delta(\vartheta-\vartheta')-
\frac1{2\pi}
\label{coscompl}\ee
for $\vartheta,\vartheta'\in[0,2\pi]$. The non-constant string modes $\xi^\mu$
can be written on $\partial\Sigma$ in
terms of the Fourier series expansion for periodic functions on the circle,
\be
\xi^\mu(\vartheta)=\sum_{n=1}^\infty\Bigl(a_n^\mu\cos n\vartheta+b_n^\mu
\sin n\vartheta\Bigr) \ .
\label{ximodeexp}\ee
The low-energy string effective action is then given by the renormalized value
of (\ref{ZFpathint}).\footnote{\baselineskip=12pt It is a curious property of
the Polyakov path integral that it computes directly the vacuum energy. The
reason becomes clearer in the effective action approach \cite{ACNY} whereby
conformal invariance is used to derive the variational equations of a spacetime
effective action for the background fields. The string partition function is
quite different from that of quantum field theory, in that it is more like an
S-matrix.}

To evaluate the path integral (\ref{ZFpathint}), we use Lorentz invariance to
rotate to a basis in which the antisymmetric $d\times d$ matrix
$F_{\mu\nu}(x_0)$ is skew-diagonal with skew-eigenvalues $f_\ell$,
$\ell=1,\dots,\frac d2$. Then, on substituting (\ref{Sboundary}),
(\ref{Ninverse}) and (\ref{ximodeexp}) into (\ref{ZFpathint}), the path
integral
factorizes into a product of $\frac d2$ functional Gaussian integrations over
the pairs of coordinate modes $a_n^{2\ell-1},a_n^{2\ell}$ and
$b_n^{2\ell-1},b_n^{2\ell}$. The result of this integration yields a functional
determinant and gives
\be
Z_0[F]=\frac1{g_s}\,\int d\vec x_0~\prod_{\ell=1}^{d/2}Z_{2\ell-1,2\ell}
[f_\ell]
\label{ZFprodfi}\ee
where
\be
Z_{2\ell-1,2\ell}[f_\ell]=\prod_{n=1}^\infty
\left(\frac{4\pi^2\alpha'}n\right)^2
\left[1+(2\pi\alpha'ef_\ell)^2\right]^{-1} \ .
\label{Z2i2i-1}\ee
The divergent infinite product $\prod_n\frac1{n^2}$ in (\ref{Z2i2i-1}) can be
regulated using the ultraviolet cutoff $\varepsilon$ and may be absorbed into a
renormalization of the string coupling constant by using zeta-function
regularization \cite{FT}. The other factor in (\ref{Z2i2i-1}) is also finite in
zeta-function regularization
\be
\prod_{n=1}^\infty c=c^{\,\zeta(0)} \ ,
\label{zetafnreg}\ee
where $\zeta(s)=\sum_n\frac1{n^s}$ is the Riemann zeta-function with
$\zeta(0)=-\frac12$. We thereby find that
$Z_{2\ell-1,2\ell}[f_\ell]=\frac1{4\pi^2\alpha'}\,
\sqrt{1+(2\pi\alpha'ef_\ell)^2}$. By
rotating back to general form, the regularized partition function
(\ref{ZFprodfi}) can be written in a Lorentz invariant way and it leads to the
effective string action
\bea
S_{\rm BI}&=&\frac1{(4\pi^2\alpha')^{d/2}g_s}\non&&\times\,
\int d\vec x_0~\sqrt{\det_{1\leq\mu,\nu\leq d}\left[\delta_{\mu\nu}
+2\pi\alpha'eF_{\mu\nu}\right]} \ ,
\label{BIaction}\eea
which describes a model of non-linear electrodynamics for the field strengths
that is governed by the classic Born-Infeld Lagrangian \cite{BI}.

The Euler-Lagrange equations for the action (\ref{BIaction}) can be written as
\be
\sqrt{\det(\id+2\pi\alpha'eF)}\,\left(\frac1{\id-(2\pi\alpha'eF)^2}
\right)_{\mu\nu}\,\beta_A^\nu=0
\label{ELBI}\ee
where
\bea
\beta_A^\mu&=&\frac\partial{\partial\ln\varepsilon}\,\delta A^\mu(\varepsilon)
\non&=&2\pi\alpha'e\,\partial_\nu F^\mu_{~\lambda}\left(\frac1
{\id-(2\pi\alpha'eF)^2}\right)^{\lambda\nu}
\label{betaA}\eea
is the one-loop worldsheet $\beta$-function \cite{ACNY,CLNY2}, with $\delta
A^\mu(\varepsilon)$ the cutoff dependent gauge field correction term that
multiplies $\dot x^\mu$ in (\ref{Sdisc}). The equations of
motion for the gauge field are therefore equivalent to worldsheet conformal
invariance of the quantum string theory. They are the stringy ${\cal
O}(\alpha')$ corrections to the Maxwell field equations for $A_\mu$. Notice
that since
$\det(\id+2\pi\alpha'eF)=\det(\id+2\pi\alpha'eF)^\top=\det(\id-2\pi\alpha'eF)$,
the Born-Infeld action (\ref{BIaction}) contains only even powers of the field
strength $F$ in an $\alpha'$ expansion. The leading order term (the field
theory limit) is given by the Maxwell Lagrangian
$-\frac{e^2T_s^{d/2-1}}{4(2\pi)^{d/2}g_s}\,F_{\mu\nu}^2$. For the uniform
electromagnetic backgrounds that we shall deal with in most of this article,
the calculations will thereby produce on-shell string amplitudes.

The Born-Infeld action is an example whereby the contributions in the coupling
constant $\alpha'$, representing the string corrections to the field theory
limit, can be summed to all orders of $\sigma$-model perturbation theory.
Born-Infeld theory has many novel characteristics which distinguish it from the
classical Maxwell theory of electromagnetism. These novel features are
predominant for a purely electric background field, which in Minkowski space
would have only non-vanishing temporal components $F_{0j}=iE_j$. Then, the
electric field generated by a point-like charge is regular at the source and
its total energy is finite \cite{BI}. The effective distribution of the field
has a radius of the order of the string length scale $\sqrt{\alpha'}$, and the
delta-function singularity is smeared away. This is quite unlike the situation
in Maxwell theory, whereby the field of a point source is singular at the
origin and its energy is infinite. The analogy with open string theory has been
used to suggest that the terms of higher order in $\alpha'$ in the string
effective action may eliminate Schwarzschild black hole singularities.
Furthermore, at the origin of the source the electric field takes on its
maximum
value $|\vec E|=E_c$. The Born-Infeld Lagrangian in this case takes the form
$\sqrt{1-(2\pi\alpha'e\vec E\,)^2}$, which shows that there is a limiting value
$E_c=T_s/e$ such that for $|\vec E|>E_c$ the action becomes complex-valued and
ceases to make physical sense \cite{FT,Burgess,Nester}. This instability
reflects the fact that the electromagnetic coupling of strings is not minimal
\cite{AN} and creates a divergence due to the fast rising density of string
states. For field strengths larger than the critical electric field value
$E_c$, the string tension $T_s$ can no longer hold the strings together. We
shall encounter other novel aspects of strings in background electric fields
throughout this paper. Notice, however, that such novel effects and
instabilities do not arise in
purely magnetic backgrounds.

Going back to the case of Euclidean signature, this calculation may be extended
to the next order in string perturbation
theory, whose contribution is the annulus diagram $\Sigma$ which by scale
invariance can be taken to have outer radius 1 and inner radius $a=\e^{-\pi
t}\in[0,1]$. The variable $a$ is therefore the modulus of the annulus and the
path integration in (\ref{calZF}) over metrics $g_{ab}$ on $\Sigma$ reduces,
after gauge fixing, to an integral over Teichm\"uller space. We may now couple
one endpoint of an open string to the boundary at $r=a$ with a charge $e_1$,
and the other end at $r=1$ with charge $e_2$, so that the one-loop action in
the conformal gauge reads
\bea
&&S_1[x,A]=\frac1{4\pi\alpha'}\,\int d^2z~\partial x^\mu\,\overline{\partial}
x_\mu\label{Sann}\\&&+\left.ie_2\int\limits_0^{2\pi}d\vartheta~\dot x^\mu
\,A_\mu(x)\right|^{r=1}-\left.ie_1
\int\limits_0^{2\pi}d\vartheta~\dot x^\mu\,A_\mu(x)\right|^{r=a} \ . \nonumber
\eea
Here we shall consider only the case of neutral strings,
$e_1=e_2=e$.\footnote{\baselineskip=12pt Because of the orientation reversal
between the two string endpoints, the net charge of an oriented open string is
$e_1-e_2$.} Charged strings will be dealt with later on.

Again, by using the method of images the Neumann function
on the annulus is found to be given by the infinite series
\bea
N(z,z')&=&\frac1{2\pi}\,\Biggl[~\ln|z-z'|\Biggr.\label{Neumannann}
\\&&+\,\sum_{n=1}^\infty
\ln\left|1-\frac{a^{2n}}{z\overline{z}'}\right|\Biggl|1-a^{2n-2}\,z
\overline{z}'\Biggr|\non&&+\left.\sum_{n=1}^\infty\ln\left|1-\frac{a^{2n}\,z}
{z'}\right|\left|1-\frac{a^{2n}\,z'}z\right|~\right] \ , \nonumber
\eea
which satisfies the usual equation of motion and the Neumann boundary
conditions $\partial_rN(z,z')|_{r=a}=0$,
$\partial_rN(z,z')|_{r=1}=\frac1{2\pi}$ \cite{ACNY}. At the worldsheet
boundaries where $z_k=\e^{i\vartheta_k}$, $k=1,2$, the annulus Green's function
(\ref{Neumannann}) can be written as
\be
N(\vartheta_k,\vartheta_l')=-\frac1\pi\,\sum_{n=1}^\infty\frac{[G_n]_{kl}}n\,
\cos n\left(\vartheta_k-\vartheta_l'\right) \ ,
\label{Nannbdry}\ee
where $k,l=1,2$ and $G_n$ is the $2\times2$ matrix
\be
G_n=\pmatrix{A_n&B_n\cr B_n&A_n\cr}
\label{Gnmatrix}\ee
with
\be
A_n=\frac{1+a^{2n}}{1-a^{2n}} \ , ~~ B_n=\frac{2a^n}{1-a^{2n}} \ .
\label{AnBndef}\ee
The function (\ref{Nannbdry}) is easy to invert and proceeding as before the
one-loop effective action may thereby be calculated to be \cite{FT,ACNY}
\be
Z_1[F]=Z_1[0]~\int d\vec x_0~
\det_{1\leq\mu,\nu\leq d}\left[\delta_{\mu\nu}+2\pi\alpha'eF_{\mu\nu}
\right] \ ,
\label{Z1F}\ee
where
\be
Z_1[0]=\frac{g_s}2\,\int\limits_0^\infty\frac{dt}t~
\left(4\pi^2\alpha't\right)^{-13}~\eta\left(\frac{it}2\right)^{-24}
\label{annvacuum}\ee
is the usual zero field vacuum energy for the annulus in the bosonic critical
dimension $d=26$, and $\eta(\tau)$ is the Dedekind function. The partition
function (\ref{annvacuum}) contains the contribution from the two conformal
ghost fields which do not couple to the external field $F_{\mu\nu}$.

This result may be straightforwardly extended to fermionic strings by using the
usual coupling of a spinor particle to an electromagnetic field and by using
anti-periodic Fourier series expansions on $\partial\Sigma$ for the fermion
fields and the corresponding string propagator. At tree level, the only effect
of supersymmetry is to cancel the tachyonic divergence that arises in
(\ref{Z2i2i-1}) \cite{Tseytrev}. The final result is again the Born-Infeld
action (\ref{BIaction}). The extension to non-abelian gauge fields is also
straightforward \cite{TseytBI} and yields the effective non-abelian Born-Infeld
action for open strings whose endpoints transform in the fundamental
representation of the gauge group. The leading term in the $\alpha'$ expansion
is the usual Yang-Mills Lagrangian for the non-abelian gauge field. Demanding
spacetime supersymmetry then leads to the usual low-energy effective field
theory description in terms of maximally supersymmetric Yang-Mills theory in
$d=10$ spacetime dimensions (the superstring critical dimension).

\subsection{Open string spectrum}

In this subsection we will describe the spectrum of open strings in a constant
background electromagnetic field in second quantization using the operator
formalism \cite{ACNY}. We will concentrate again on bosonic strings, as we are
merely interested here in some of the basic qualitative features of the
spectrum. We will assume that the string worldsheet $\Sigma$ is now an infinite
strip with coordinates $(\tau,\sigma)$, where $\tau\in\real$ and
$\sigma\in[0,1]$ (This surface is conformally equivalent to the disc). The
Euclidean action is given by
\bea
S_{\rm strip}&=&\frac 1{4\pi\alpha^\prime}\,\int d\tau~d\sigma~
\left( \d_\tau x^\mu\,\d_\tau x_\mu+\d_\sigma x^\mu\,\d_\sigma x_\mu \right)
\non&&+\left.\frac{e_2}2\,\int d\tau~F_{\mu\nu}\,x^\nu\,\partial_\tau x^\mu
\right|^{\sigma=1}\non&&
-\left.\frac{e_1}2\,\int d\tau~F_{\mu\nu}\,x^\nu\,\partial_\tau x^\mu
\right|^{\sigma=0} \ ,
\label{Sstrip}\eea
and in the worldsheet canonical formalism we regard $\tau$ as the time
coordinate and $\sigma$ as the space coordinate (so that $\Sigma$ now has
Minkowski signature). Varying (\ref{Sstrip}) gives the usual wave equation
$\Box\,x^\mu=0$ along with the mixed Neumann-Dirichlet boundary conditions
\bea
\left(\partial_\sigma x^\mu-2\pi\alpha'e_1F^\mu_{~\nu}\,\partial_\tau x^\nu
\right)_{\sigma=0}&=&0 \ , \non\left(\partial_\sigma x^\mu-2\pi\alpha'e_2
F^\mu_{~\nu}\,\partial_\tau x^\nu\right)_{\sigma=1}&=&0 \ .
\label{NDbcs}\eea
We will again use Lorentz invariance to skew diagonalize the real-valued
antisymmetric tensor $F_{\mu\nu}$. Since the skew blocks are independent, it
suffices to concentrate on only one of them, and so we assume that the only
non-vanishing component of the field strength tensor is $F_{01}=-F_{10}=F$. In
this plane of the field, we introduce the complex target space coordinates
$x^\pm=\frac1{\sqrt2}(x^0\pm ix^1)$, in terms of which the boundary conditions
(\ref{NDbcs}) become
\bea
\left(\partial_\sigma x^++2\pi i\alpha'e_1F\,\partial_\tau x^+
\right)_{\sigma=0}&=&0 \ , \non\left(\partial_\sigma x^-+2\pi i\alpha'e_2
F\,\partial_\tau x^-\right)_{\sigma=1}&=&0 \ ,
\label{pmbcs}\eea
along with the standard free open string Neumann boundary conditions in all of
the directions transverse to the 0--1 plane.

We will now write down mode expansions which solve the equations of motion and
satisfy the requisite boundary conditions (\ref{pmbcs}). Since the only
modification from the usual free string case occurs for the harmonic string
coordinates in the 0--1 plane, we will focus our attention on their
contributions. For this, it is necessary to treat neutral and charged open
strings separately. As we will see in the following, there are drastic
differences at both a qualitative and analytic level between the two cases. Let
us note, however, that unitarity of the open string theory always requires the
existence of both charged and neutral strings in the spectrum \cite{AMSS}.
Consider a string scattering amplitude with the given charges $e_1,e_2$ at the
endpoints. An amplitude with an even number of external legs can be sliced in
many different ways into intermediate states. Some of these intermediate states
will consist of open strings with either the charge $e_1$ or $e_2$ at both of
its ends. An amplitude with an odd number of external legs necessarily involves
at least one neutral string state in the scattering process. Therefore, any
amplitude should be summed over {\it all} charges in the decomposition of the
fundamental representation of the Chan-Paton gauge group under the embedding of
$U(1)$ induced by the background electromagnetic field.

\subsubsection{Neutral strings}

Let us begin with the case where the total charge of the open string vanishes,
$e_1=e_2=e$. In this case there is the freedom to add to the coordinates
$x^\pm$ terms proportional to $\tau\mp2\pi^2i\alpha'e\sigma$, which satisfy the
boundary conditions (\ref{pmbcs}) when $e_1-e_2=0$. The mode expansions in the
0--1 plane can thereby be written as
\bea
&&x^\pm(\tau,\sigma)=\frac{y^\pm+q_\mp\left[\tau\mp2\pi^2i\alpha'eF
(\sigma-\frac12)\right]}{\sqrt{1+(2\pi\alpha'eF)^2}}\non&&+\,i\,
\sum_{n=1}^\infty\left(\frac{a_n^\pm}n~\e^{-in\tau}\,\cos\left(n\pi\sigma\pm
\arctan2\pi\alpha'eF\right)\right.\non&&-\left.\frac{(a_n^\mp)^\dagger}n~
\e^{in\tau}\,\cos\left(n\pi\sigma\mp\arctan2\pi\alpha'eF\right)\right) \ ,
\non&&
\label{xpmneutralmodes}\eea
with $(y^\pm)^\dagger=y^\mp$ and $(q_\pm)^\dagger=q_\mp$. The expansions
(\ref{xpmneutralmodes}) are defined in terms of an orthonormal system of
oscillation modes \cite{ACNY,Burgess,Nester} which solves the variational
problem for the action (\ref{Sann}) on the infinite strip and which
diagonalizes it. The canonical momenta conjugate to the fields $x^\pm$ are
given by $p_\pm=\partial_\tau x^\pm$ and they lead to the canonical commutation
relations of the quantum string theory in the usual way. Because of the
Born-Infeld factor in the denominator of the first line of
(\ref{xpmneutralmodes}), one finds that the zero mode positions $y^\pm$ and
momenta $q_\pm$ obey canonical (Heisenberg) commutation relations,
respectively, and are mutually commutative otherwise. The Fourier modes obey
the Heisenberg-Weyl algebra $[a_m^\pm,(a_n^\mp)^\dagger]=n\,\delta_{nm}$. They
therefore satisfy the same harmonic oscillator commutation relations that they
would in the absence of the external field.

The Hamiltonian density is $(\partial_\tau x^++\partial_\sigma
x^+)(\partial_\tau x^-+\partial_\sigma x^-)$ which leads to the total
worldsheet Hamiltonian
\be
L_0^\parallel=q_+q_-+\sum_{n=1}^\infty\left((a_n^-)^\dagger\,a_n^+
+(a_n^+)^\dagger\,a_n^-\right)
\label{L0neutral}\ee
in the 0--1 plane, with $q_+q_-=\frac12\,(q_1^2+q_2^2)$. We conclude that the
spectrum of a neutral open string is not affected by the electromagnetic field.
However, as we saw in the previous subsection, the vacuum-to-vacuum amplitude
is modified because the usual Born-Infeld factor appears in the mass-shell
condition \cite{ACNY}.

\subsubsection{Charged strings}

When the total charge of the string is non-vanishing, the entire structure of
the external field problem is different. The string fields no longer have
integer oscillator modes, and the zero modes change completely. In particular,
there is no function linear in $\tau$ and $\sigma$ which can satisfy the
boundary conditions (\ref{pmbcs}) when $e_1-e_2\neq0$. The mode expansion in
this case is given by
\bea
x^\pm(\tau,\sigma)&=&y^\pm\mp i\,a_0^\pm\,\frac{\e^{\pm i\alpha\tau}}\alpha
\label{xpmchargedmodes}\\&&\times\,\cos\left(\pi\alpha\sigma-
\arctan2\pi\alpha'e_1F\right)\non&&+\,i\,
\sum_{n=1}^\infty\left[\frac{a_n^\pm}{n\mp\alpha}~\e^{-i(n\mp\alpha)\tau}
\right.\non&&\times\,\cos\Bigl((n\mp\alpha)\pi\sigma\pm\arctan2\pi\alpha'e_1F
\Bigr)\non&&-\,\frac{(a_n^\mp)^\dagger}{n\pm\alpha}~
\e^{i(n\pm\alpha)\tau}\non&&
\times\Biggl.\cos\Bigl((n\pm\alpha)\pi\sigma\mp\arctan2\pi\alpha'e_1F
\Bigr)\Biggr] \ , \nonumber
\eea
where
\be
\alpha=\frac1\pi\,\Bigl(\arctan2\pi\alpha'e_1F-\arctan2\pi\alpha'e_2F\Bigr) \ .
\label{alphacharged}\ee
We will assume in this subsection that $e_1>e_2$ so that $\alpha>0$. The normal
mode functions in (\ref{xpmchargedmodes}) again diagonalize the action
(\ref{Sann}), and solve the wave equation and the boundary conditions
(\ref{pmbcs}) \cite{ACNY,Burgess,Nester}. Note that since their integrals are
non-zero, the $y^\pm$ cannot be identified with the center of mass coordinates
of the open strings, in contrast to the neutral case. Notice also the
appearence of the extra modes $a_0^\pm$ compared to the $\alpha=0$ case, which
by
reality are required to be Hermitian operators. Canonical quantization now
identifies the quantum commutators
\bea
\left[a_n^\pm\,,\,(a_m^\mp)^\dagger\right]&=&(n\pm\alpha)\,\delta_{nm} \ ,
\label{anpmcomm}\\\left[y^+\,,\,y^-\right]&=&\frac1{2\alpha'F}\,
\frac1{e_1-e_2} \ .
\label{ypmcomm}\eea
The drastic change in the zero mode structure for charged strings is apparent
in the commutation relation (\ref{ypmcomm}), which is ill-defined in the
neutral string limit $e_1=e_2$.

The total worldsheet Hamiltonian in the 0--1 plane can be worked out to be
\be
L_0^\parallel=\sum_{n=1}^\infty(a_n^+)^\dagger\,a_n^-+
\sum_{n=0}^\infty(a_n^-)^\dagger\,a_n^++\frac12\,\alpha(1-\alpha) \ .
\label{L0charged}\ee
The normal ordering constant in (\ref{L0charged}) depends on the (arbitrary)
choice of $a_0^+$ as an annihilation operator and is required to put the
Virasoro algebra in standard form \cite{ACNY,Nester}. We see therefore that the
external electromagnetic field has a drastic effect on the spectrum of charged
strings. It shifts the oscillation frequencies by amounts $\pm\alpha$, it
modifies the commutation relations of the zero modes, and it changes the zero
point energy. Furthermore, the open string momentum operators $q_\pm$ no longer
appear in the mode expansions, while there are extra Fourier operators
$a_0^\pm$ which create and annihilate quanta of frequency $\alpha$. In fact,
the contribution from the coordinates in the 0--1 plane is formally identical
to that of a twisted unprojected sector of an orbifold string theory with twist
angle $\alpha$. This orbifold analogy provides a computationally convenient
characterization of the external field problem.

Normally, one would take the quantum states to be eigenstates of definite
momentum. However, when $e_1\neq e_2$, it is instead the zero mode operators
$y^\pm$ that commute with the Hamiltonian $L_0^\parallel$, and so we may take
the states to be eigenstates of $y^+$, for example. Note that the operator
$y^-$ is, according to (\ref{ypmcomm}), a conjugate momentum operator for
$y^+$. Since $L_0^\parallel$ does not depend on $y^\pm$, there is an infinite
degeneracy in the spectrum. In fact, the present physical situation is
identical to that of a charged particle moving in the plane under the influence
of a perpendicular uniform magnetic field. The states form equally spaced
Landau levels of infinite degeneracy, with the energy difference between
consecutive levels proportional to $(e_1-e_2)F$. The operators $a_0^\pm$ move
the string from one Landau level to another, and their frequency separation
(\ref{alphacharged}) is proportional to the quantity $(e_1-e_2)F$ when it is
sufficiently small, i.e. in the weak-field limit
$\alpha=2\alpha'(e_1-e_2)F+{\cal O}(F^3)\ll1$. Deviations from the field
theoretic result at strong fields $F$ come from the non-minimality of the
electromagnetic string coupling and are parametrized by the non-linear function
$\alpha$ of the field. Excited states of the open string are then obtained by
acting on the ground state wavefunctions with oscillator creation operators. At
the first excited level, there are the states $(a_1^+)^\dagger|y^+\rangle$ with
tachyonic mass $\sqrt{-\frac12\,\alpha(1+\alpha)}$, and the states
$(a_1^-)^\dagger|y^+\rangle$ with mass $\sqrt{\frac12\,\alpha(3+\alpha)}$. This
is reminescent of the situation that occurs in Yang-Mills theory in the
presence of a chromomagnetic field condensate, whereby one gluonic polarization
becomes unstable due to its tachyonic energy and the other becomes massive
\cite{NO}.\footnote{\baselineskip=12pt In contrast to the usual bosonic string
tachyon state, this instability is not removed by supersymmetry \cite{FT1}.} In
fact, as we shall see in the following, the charged string system possesses
many instabilities. Only the neutral open string makes sense both physically
and analytically.

\subsection{The Schwinger mechanism}

In this subsection we will compute the one-loop vacuum energy for charged
strings using the operator formalism of the previous subsection and elucidate
somewhat on the instabilities that we have thus far encountered. In particular,
we will examine the instability of the string vacuum in a purely electric
background \cite{Burgess,BP}. This can be obtained from the calculations of the
previous subsection by the analytical continuations $F=iE$ and
$\alpha=-i\epsilon$ corresponding to Wick rotations of both the worldsheet and
target space time coordinate to Minkowski signature. The vacuum energy may be
computed as the logarithm of the partition function $\det(L_0-1)^{-1/2}$ of the
underlying free conformal field theory $\sigma$-model, where
$L_0=L_0^\perp+L_0^\parallel+L_0^{\rm ghosts}$ is the total Hamiltonian
comprised of the contributions from the fields transverse to the plane of the
electric field, those parallel to the 0--1 plane, and the conformal ghost
fields. The annulus amplitude is thereby given as
\be
-i\,V_d\,{\cal F}(e_1,e_2)=\frac12\,\tr_{(e_1,e_2)}\ln\left(L_0-1\right) \ ,
\label{annamplcharged}\ee
where $V_d$ is the volume of spacetime and $\tr_{(e_1,e_2)}$ denotes the trace
over all string states in the $(e_1,e_2)$ charge sector. The total annulus
amplitude is a sum over all allowed endpoint charges.

The trace (\ref{annamplcharged}) is straightforward to evaluate by using the
proper time representation
\be
\ln{\cal A}=-\int\limits_0^\infty\frac{dt}t~\e^{-\pi t{\cal A}} \ ,
\label{lncalDint}\ee
and the fact that for any set of oscillator operators $a_n$ obeying
Heisenberg-Weyl commutation relations there is the formula
\bea
&&\tr~\e^{-\pi t\sum_{n\geq1}a_n^\dagger\,a_n}=\prod_{n=1}^\infty\tr~
\e^{-\pi t\,a_n^\dagger\,a_n}\label{osctrace}\\
&&~~=\prod_{n=1}^\infty~\sum_{m=0}^\infty
\e^{-\pi tmn}=\prod_{n=1}^\infty\left(1-\e^{-\pi tn}\right)^{-1} \ , \nonumber
\eea
where we have used a basis of all possible multi-particle states. For the
transverse degrees of freedom the oscillator traces are accompanied by $d-2$
Gaussian momentum integrals, coming from the analogs of the first term in
(\ref{L0neutral}), and an integration over the canonically conjugate zero modes
$y^\perp$ of the string fields which produces a volume factor $V_{d-2}$.
These are also multiplied by a factor $(2\pi)^{d-2}$ which is the density of
quantum phase space states. For the fields along the 0--1 plane, we can use the
identity (\ref{osctrace}) with the appropriate shifts of oscillation
frequencies given in (\ref{anpmcomm}). There is also an integration over the
zero modes $y^\pm$ which contribute, according to (\ref{ypmcomm}), a quantum
state density factor $\frac{\alpha'}\pi\,E\,(e_1-e_2)$, along with the volume
of the 0--1 plane.

By incorporating the ghost contributions and putting all of
these results together we arrive finally at
\bea
{\cal F}(e_1,e_2)&=&\frac12\,\int\limits_0^\infty\frac{dt}t~
\left(4\pi^2\alpha't\right)^{-13}\,\eta\left(\frac{it}2\right)^{-24}
\non&&\times\,C_A(t,E)
\label{annamplfinal}\eea
in the critical dimension $d=26$, where
\be
C_A(t,E)=\alpha'(e_1-e_2)E\,t~\e^{-\pi t\epsilon^2/2}~
\frac{\Theta^\prime_1\left(0\left\vert
\frac{it}2\right.\right)}{\Theta_1\left(\frac{\epsilon t}2\left\vert
\frac{it}2\right.\right)}
\label{fAtE}\ee
is a field dependent correction factor. Here $\Theta_a(\nu|\tau)$ denote the
standard Jacobi-Erderlyi theta-functions and
$\Theta_1'(\nu|\tau)=\frac\d{\d\nu}\,\Theta_1(\nu|\tau)$. In the zero field
limit the amplitude (\ref{annamplfinal}) reduces to the expected result
(\ref{annvacuum}), since $C_A(t,0)=1$. It gives the modification of the neutral
string effective action (\ref{Z1F}) to the charged case. In the limit
$e_1-e_2=\delta\to0$, we have
$\pi\epsilon=\frac\delta{1-(2\pi\alpha'e_1E)^2}+{\cal O}(\delta^2)$, and the
correction factor (\ref{fAtE}) takes the simple $t$-independent form
$C_A(t,E)=1-(2\pi\alpha'e_1E)^2+{\cal O}(\delta)$. Thus, for neutral strings
the annulus amplitude (\ref{annamplfinal}) is proportional to the square of the
Born-Infeld Lagrangian for this case, as in (\ref{Z1F}).

The most interesting feature of the vacuum energy (\ref{annamplfinal}) is that
it is imaginary. The theta-function appearing in the denominator of
(\ref{fAtE}) contains a trigonometric function, $\Theta_1(\frac{\epsilon
t}2|\frac{it}2)\propto\sin(\frac{\pi t\epsilon}2)$, and so the function
$C_A(t,E)$ has simple poles on the positive $t$-axis at $t=2k/|\epsilon|$,
$k=1,2,\dots$. The amplitude thereby acquires an imaginary part given by the
sum of the residues at the poles times a factor of $\pi$, since, as dictated by
the proper definition of the Feynman propagator, the contour of integration
in the complex $t$-plane should pass to the right of all poles. What this
quantum instability represents is the spontaneous creation of charged strings
from the vacuum \cite{Burgess,BP}, in analogy to the instability of the vacuum
state in quantum electrodynamics \cite{Schwinger}. By computing the
corresponding residues of the function (\ref{fAtE}), the total rate of pair
production is found to be given by
\bea
w_{\rm str}&=&-2\,{\rm Im}\,{\cal F}\non&=&\frac1{(2\pi)^d}\,\sum_{e_1,e_2}
\,\sum_S\frac{\alpha'(e_1-e_2)}\epsilon\label{wstr}\\&&\times\,
\sum_{k=1}^\infty(-1)^{k+1}\,\left(\frac{|\epsilon|}k\right)^{d/2}~
\e^{-\frac{2\pi k}{|\epsilon|}\,\bigl(M_S(\epsilon)^2+\frac{\epsilon^2}2
\bigr)} \ , \nonumber
\eea
where the second sum runs through all physical open string states of mass
$M_S(\epsilon)$ which may be computed from the generating function
\bea
&&\frac{\e^{\bigl(1-\frac{\epsilon^2}2\bigr)l}}{\sin\left(\frac{\epsilon l}2
\right)}\,\prod_{n=1}^\infty\frac{(1-\e^{-nl})^{-(d-4)}}
{(1-\e^{-(n+i\epsilon)l})
(1-\e^{-(n-i\epsilon)l})}\non&&~~=\sum_S\e^{-M_S(\epsilon)^2\,l} \ .
\label{MSgenfn}\eea
Note that neutral string states do not contribute to the pair production rate,
as expected, and indeed the neutral string vacuum energy
(\ref{Z1F},\ref{annvacuum}) is real-valued.

The expression (\ref{wstr}) represents the stringy modification of the classic
Schwinger probability amplitude for pair creation of charged particles in a
uniform external electric field $E$ \cite{Schwinger}. In that case the
probability per unit volume and unit time is given by
\bea
w_0&=&\frac{2J+1}{(2\pi)^3}\,\sum_{k=1}^\infty(-1)^{(2J+1)(k+1)}\non&&\times\,
\left(\frac{QE}k\right)^2~\e^{-2\pi kM^2/|QE|} \ ,
\label{Schwinger}\eea
where $Q$, $J$ and $M$ are the charge, spin and mass of the created
particles. In this quantum field theory calculation the imaginary part of the
vacuum energy comes from the determinant $\det(-D_A^2)^{-1/2}$ of the massive
gauge-covariant Dirac operator $D_A$, which at tree-level would produce the
result $\frac{\pi M^{-1}QE}{\sin(\pi M^{-1}QE)}$. In fact, the result
(\ref{wstr}) coincides with (\ref{Schwinger}) with $Q=2\alpha'(e_1-e_2)$ in the
weak-field limit in $d=4$ dimensions, since a particle-antiparticle pair of
spin $J$ has $2(2J+1)$ physical states.

However, in contrast to the field
theory case, the string theory deteriorates at strong external fields. Since
$\epsilon\to\infty$ as $E\to E_c=T_s/e_1$, the total rate for pair production
diverges at the critical electric field \cite{BP}. Thus the classical
instability of the string vacuum state in an electric field can also be seen at
the quantum level. At this critical value of the external field, the string
tension can no longer stop charged strings from nucleating out of the vacuum.
In fact, this limiting instability also occurs for neutral strings. If we
concentrate on only the first line of (\ref{xpmneutralmodes}) (the zero mode
contributions), then we see that the open string can be thought of as a rod, of
length proportional to $E\,q$, which behaves like an electric dipole whose ends
carry equal and opposite charges. When an open string is stretched along the
direction of the background electric field, the field reduces its energy, and
at $E=E_c$ the energy stored in the tension of the string is balanced by the
electric energy of the stretched string. For $E>E_c$, virtual strings can
materialize out of the vacuum, stretch to infinity and destabilize the ground
state. In fact, from the first line of (\ref{xpmneutralmodes}) we see that for
fixed worldsheet time $\tau$ the two endpoints of the string are not at the
same value of $x^+$, but they are always spacelike separated. As the electric
field becomes critical the two ends at fixed $\tau$ become lightlike separated.

Of course, in the genuine Type I theory that these considerations really
pertain to, one should add to the annulus amplitude the contribution from its
non-orientable counterpart, the M\"obius strip diagram. This is straightforward
to do and the role of the M\"obius amplitude is to project out the
reflection-odd, neutral oriented string states
\cite{BP}.\footnote{\baselineskip=12pt Since the M\"obius strip has only a
single connected boundary, only neutral string states contribute to the
M\"obius amplitude. Physical string states must also be even under a worldsheet
parity reflection of the open strings.} One should also, for unitarity reasons,
consider the contributions from the one-loop closed string diagrams, i.e. the
torus and the Klein bottle. Since closed string states do not couple to the
external field, their amplitudes are the same as in the zero field limit. The
four contributions to the total vacuum energy now have an elegant
interpretation in terms of the worldsheet orbifold construction that we
mentioned earlier, whereby the torus and annulus diagrams give the
contributions of untwisted and twisted sectors, respectively, while the
addition of the Klein bottle and M\"obius strip diagrams takes care of the
projections onto states which are even under the action of the orbifold group.
The extension of the calculation to open superstrings is also straightforward
and again one easily recovers the Schwinger formula (\ref{Schwinger}) in the
weak-field limit \cite{BP,Tseyt1,Tseytrev}.

\section{Thermal ensembles}

In this section we will describe some properties of strings in background
electromagnetic fields at finite temperature. For this, we are interested in
computing the thermodynamic partition function
\be
{\cal Z}=\tr~\e^{-\beta(L_0-1)} \ ,
\label{thermpartfn}\ee
where $\beta=1/k_{\rm B}T$ with $k_{\rm B}$ the Boltzmann constant and $T$ the
temperature. Temperature represents another explicit supersymmetry breaking
mechanism and it leads to a variety of novel effects in string theory. At the
forefront of these exotic features is the influence of the density of single
particle states on the thermodynamic properties of the string gas. The number
of states at level $N$ grows exponentially as $\e^{4\pi\sqrt{N}}$, which is so
rapid that the thermodynamic partition function (\ref{thermpartfn}) of the free
string gas converges only for sufficiently small temperatures $T<T_{\rm H}$,
where
\be
T_{\rm H}=\frac1{2\pi k_{\rm B}\sqrt{2\alpha'}}
\label{TH}\ee
is known as the Hagedorn temperature \cite{Hagedorn}. Generally, models with an
exponentially rising density of states exhibit non-extensive thermodynamic
quantities and a pair of such systems can never attain thermal equilibrium.
However, in string theory the Hagedorn temperature is not a limiting
temperature, because it requires a finite amount of energy to reach it in the
canonical ensemble. Rather, it is associated with a phase transition, analogous
to the deconfinement transition that occurs in Yang-Mills theory. The Hagedorn
temperature $T_{\rm H}$ is the critical point at which infrared divergences
emerge due to a closed string state becoming massless \cite{Kogan,Sathi,AW}.
The Hagedorn transition may therefore be associated with the appearence of
tachyonic winding modes. In the following we will examine how this picture is
affected by the presence of background fields.

Although the thermodynamic ensemble of superstrings is interesting in its own
right \cite{LBO}, the inclusion of electromagnetic fields will allow us in the
next section to map the free string gas onto a system of D-branes. Thermal
states of superstrings in electromagnetic fields thereby correspond to
non-extremal states of D-branes in supergravity which have a natural Hawking
radiation and entropy. They are therefore relevant to the microscopic
description of black holes in string theory. Since the string theory
universally contains gravity, the system destabilizes at finite energy density
in the thermodynamic limit. This is due to the Jeans instability which occurs
because a relativistic thermal ensemble at sufficiently large volume reaches
its Schwarzschild radius and collapses into a black hole \cite{AW}.

In the path integral approach to finite temperature string theory, the
spacetime is taken as Euclidean space with time $x^0$ compactified on a circle
of circumference $\beta$. Temperature affects the string gas because the string
can wrap around the compact time direction with a given winding number
$n\in\zed$, i.e. $x^0(r,\vartheta+2\pi)=x^0(r,\vartheta)+n\beta$. This affects
only
the zero modes of the bosonic string embedding field $x^0$ and can be
incorporated by adding a term $\frac{n\beta}{2\pi}\,\vartheta$ to its mode
expansion. In string perturbation theory, the disc amplitude (\ref{BIaction})
is unmodified at finite temperature, because the disc worldsheet cannot wrap
the cylindrical target space and so cannot distinguish between a compactified
and an uncompactified spacetime. The first corrections due to temperature
appear in the annulus amplitude. We can evaluate the thermodynamic free energy
${\cal F}=-\ln({\cal Z})/\beta$ as before by computing the Polyakov path
integral for the action (\ref{Sann}) and enforcing the periodicity constraint
in Euclidean time via the substitution
\be
x^0(r,\vartheta)\mapsto x^0(r,\vartheta)+\frac{n\beta}{2\pi}\,\vartheta \ .
\label{x0per}\ee
We then sum the path integral over all temperature winding modes $n\in\zed$.

For a uniform electromagnetic field, there are two commonly used gauge
choices, the static gauge $(A_0,A_i)=(-\vec F\cdot\vec x,\frac12\,F_{ij}\,x^j)$
and the temporal gauge $(A_0,A_i)=(0,F_i\,x^0+\frac12\,F_{ij}\,x^j)$. Here the
vector $\vec F$ denotes the temporal components of the Euclidean field strength
tensor, $F_i=F_{0i}$, $i=1,\dots,d-1$, which is related to the electric field
$\vec E$ in Minkowski space by $\vec F=i\vec E$. In the temporal gauge, the
gauge potential is only periodic in Euclidean time up to a gauge
transformation, and in this case it is necessary to augment the usual coupling
of the edge of a charged open string to the gauge field by adding a generalized
Wu-Yang term \cite{AMSS} in order to compensate the gauge transformation. Here
we shall choose the periodic static gauge field configuration. On substituting
(\ref{x0per}) into (\ref{Sann}) the action then reads
\bea
&&S_1[x,A]=\frac1{4\pi\alpha'}\,\int d^2z~\partial x^\mu\,\overline{\partial}
x_\mu+\frac{n^2\beta^2t}{32\pi\alpha'}\label{SannT}\\&&+\left.
\frac{ie_2n\beta}{2\pi}\,\int
\limits_0^{2\pi}d\vartheta~\vec F\cdot\vec x\,\right|^{r=1}-\left.
\frac{ie_1n\beta}{2\pi}\,
\int\limits_0^{2\pi}d\vartheta~\vec F\cdot\vec x\,\right|^{r=a}\non&&
+\left.ie_2\int\limits_0^{2\pi}d\vartheta~\dot x^0\,\vec F\cdot\vec x\,
\right|^{r=1}-\left.ie_1\int\limits_0^{2\pi}d\vartheta~\dot x^0\,\vec
F\cdot\vec x
\,\right|^{r=a}\non&&+\left.\frac{ie_2}2\,\int\limits_0^{2\pi}d\vartheta~\dot
x^i
\,F_{ij}\,x^j\right|^{r=1}-\left.\frac{ie_1}2\,\int\limits_0^{2\pi}d\vartheta~
\dot x^i\,F_{ij}\,x^j\right|^{r=a} \ . \nonumber
\eea
It is clear from (\ref{SannT}) that electric and magnetic backgrounds
contribute very differently at finite temperature, and so we shall analyse
their couplings separately.

\subsection{Magnetic fields}

Let us begin with the purely magnetic case and set $\vec F=\vec0$ in
(\ref{SannT}). Let $f_\ell$, $\ell=1,\dots,\frac{d-2}2$, be the
skew-eigenvalues of the magnetic flux tensor $F_{ij}$. The zero temperature
annulus amplitude when $F_{ij}$ has only a single non-vanishing skew-eigenvalue
$f_\ell$ is given by (\ref{annamplfinal},\ref{fAtE}) with $E=-if_\ell$ and
$\epsilon=i\alpha(f_\ell)$, and it is real-valued. Since the mode expansions of
the string fields are the same as before, it follows that the only effect of
finite temperature comes from the constant term in the first line of
(\ref{SannT}). By summing the path integral over all thermal winding modes,
this inserts into the Teichm\"uller integration defining the annulus amplitude
the infinite series
\be
\sum_{n=-\infty}^\infty\e^{-\beta^2n^2t/32\pi\alpha'}=\Theta_3\left(0\left|
\frac{i\beta^2t}{32\pi^2\alpha'}\right.\right) \ .
\label{Theta3T}\ee
The one-loop free energy per unit volume of the string gas in the magnetic
background is thereby given as
\bea
{\cal F}_{\rm mag}&=&\frac12\,\int\limits_0^\infty\frac{dt}t~\left(
4\pi^2\alpha't\right)^{-13}\,\eta\left(\frac{it}2\right)^{-24}
\non&&\times\,\Theta_3\left(0\left|\frac{i\beta^2t}{32\pi^2\alpha'}
\right.\right)\,\prod_{\ell=1}^{12}C_A(t,-if_\ell) \ .
\label{FmagT}\eea
We will now examine the convergence properties of the Teichm\"uller integral
(\ref{FmagT}).

The open string ultraviolet behaviour is determined by the $t\to0$ region of
moduli space. In this limit the theta functions appearing in
(\ref{FmagT},\ref{fAtE}) have the asymptotics $\eta(\frac{it}2)^{-24}\sim
t^{12}\,\e^{4\pi/t}$, $\Theta_1(\frac{i\alpha t}2|\frac{it}2)\sim
t^{-1/2}\,\e^{-\pi/2t}\,\e^{\pi\alpha^2t/2}\sin\pi\alpha$, and
$\Theta_3(0|\frac{i\beta^2t}{32\pi^2\alpha'})-1\sim
t^{-1/2}\,\e^{-32\pi^3\alpha'/\beta^2t}$. From these behaviours it follows that
the integral (\ref{FmagT}) converges in the region $t\to0$ provided that
$\beta>1/k_{\rm B}T_{\rm H}$, where $T_{\rm H}$ is the Hagedorn temperature
(\ref{TH}).\footnote{\baselineskip=12pt Of course, the bosonic string theory
has a tachyonic instability at any temperature. The Hagedorn temperature is
defined in this case as the temperature at which the one-loop free energy of
the bosonic string gas diverges even if the contribution from the usual tachyon
mode is ignored.} Note that the overall asymptotics are completely independent
of the external field, and we therefore conclude that the presence of a
magnetic field does not change the value of the Hagedorn temperature of the
free open string gas. The open string infrared behaviour, on the other hand,
comes from the $t\to\infty$ region of moduli space. In this limit the
temperature dependence of (\ref{FmagT}) disappears,
$\Theta_3(0|\frac{i\beta^2t}{32\pi^2\alpha'})\sim1$, and the conditions for
convergence of the integral are the same as at zero temperature. One encounters
the infrared magnetic instability that was described in section II.B.2
\cite{Tseyt1}. This instability in the thermodynamic free energy is also
present in supersymmetric Yang-Mills theory at both zero and finite
temperature.

It is straightforward to repeat the analysis for the fermionic string (where
there is no zero temperature tachyon mode). As spacetime bosons are required to
obey periodic boundary conditions along the Euclidean time circle and spacetime
fermions to obey anti-periodic ones, supersymmetry is explicitly broken and the
GSO projection must be accordingly modified. At the one-loop level this may be
achieved by inserting an extra weighting $(-1)^n$ in the sum over temperature
winding numbers for the $(-,+)$ spin structure in the Neveu-Schwarz sector of
the worldsheet theory \cite{AW}. Then one can compute that the instabilities
encountered above persist for superstrings \cite{Tseyt1}.

\subsection{Electric fields}

Now let us consider the case of a purely electric background and set $F_{ij}=0$
in (\ref{SannT}). From the second line of (\ref{SannT}) we see that there is
now a non-trivial coupling between the electric field and the temperature
winding modes. This coupling has several dramatic effects on the thermal
ensemble. The most glaring one is that it prevents the formation of an
equilibrium distribution of charged strings in the electric field \cite{AMSS}.
To see this, we consider the zero modes $x^\mu=x^\mu_0$ of the string embedding
fields on the annulus. In the absence of an electric field, the action is
independent of them and integrating them out in the path integral produces a
volume factor $\beta V_{d-1}$. In the present case, however, they contribute
the quantity
\bea
&&\beta\int d\vec x_0~\e^{i(e_2-e_1)n\beta\vec F\cdot\vec x_0}\non&&~~=
\beta\,(2\pi)^{d-1}\,\delta\Bigl((e_2-e_1)n\beta\vec F\,\Bigr) \ .
\label{zeromodeint}\eea
This result shows that thermal states of the string are stable only either for
neutral strings or in the absence of the external field. All states except the
ground state $n=0$ contain excitations of charged particles and therefore have
infinite energy. In fact, because of the Schwinger mechanism that we described
in section II.C, even the ground state is unstable. The breaking of
translational invariance forbids an equilibrium state of charged strings in a
constant background electric field. In what follows we shall describe some
origins of this instability.

\subsubsection{Neutral strings}

It is natural to consider the neutral string case, $e_1=e_2=e$. In that case
the zero modes $\vec x_0$ disappear from the action (\ref{SannT}), but the
second line still contributes a linear term to the Gaussian form for the
oscillatory modes $\vec\xi(\vartheta)$ of the string fields. On completing the
square, this adds an extra term $\beta^2n^2t\alpha'e^2E^2/8$ to the argument of
the exponential in the infinite series (\ref{Theta3T}), where $E=-i|\vec F\,|$.
The modification of the annulus amplitude (\ref{Z1F},\ref{annvacuum}) to finite
temperature is therefore the free energy
\bea
{\cal F}_{\rm el}&=&\frac12\,\Bigl(1-(2\pi\alpha'eE)^2\Bigr)
\int\limits_0^\infty
\frac{dt}t~\left(4\pi^2\alpha't\right)^{-13}\label{Fel}\\&&\times\,
\eta\left(\frac{it}2\right)^{-24}\,\Theta_3\left(0\left|\frac{i\beta^2t
\left[1-(2\pi\alpha'eE)^2\right]}{32\pi^2\alpha'}\right.\right) \ . \nonumber
\eea
By repeating the asymptotic analysis of the previous subsection, we find that
the Teichm\"uller integral (\ref{Fel}) converges in the open string ultraviolet
regime provided that $\beta>1/k_{\rm B}T_{\rm H}(E)$, where
\be
T_{\rm H}(E)=\frac{\sqrt{1-(2\pi\alpha'eE)^2}}{2\pi k_{\rm B}
\sqrt{2\alpha'}} \ .
\label{THE}\ee
Thus, in contrast to the magnetic case, an electric background modifies the
Hagedorn temperature (\ref{TH}) of the neutral string gas by the familiar
Born-Infeld Lagrangian \cite{FFdela}.

This result was to be expected because, unlike magnetic fields, electric fields
couple to the temporal coordinate and therefore scale the momentum of the
strings. In turn, they rescale the proper time variable $t$. Note that the
Hagedorn temperature (\ref{THE}) decreases with increasing electric field and
vanishes at the critical value $E=E_c$. As is apparent in second quantization
\cite{Nester,FFdela} (see Eq.~(\ref{xpmneutralmodes})), the field dependent
rescaling factor originates as a modification of the string tension
$T_s=1/2\pi\alpha'$, which determines the critical temperature (\ref{TH}), to
an effective tension
\be
T_{\rm eff}=T_s\left[1-(T_s^{-1}eE)^2\right]
\label{Teff}\ee
which vanishes at the critical electric field. This modification of the tension
is the reason why the thermodynamic properties of the neutral string gas are
altered by the electric background. Again the same conclusions are reached for
the full superstring free energy \cite{Tseyt1}.

\subsubsection{Charged strings}

Restricting the spectrum to neutral open strings does not completely cure the
electric field instability, because we have to sum over all allowed neutral and
charged string states. We will now examine the reasons why finite temperature
string theory forbids constant electric fields. This instability can in fact be
seen at the field theoretical level. The coordinate space diagonal elements of
the (un-normalized) thermal density matrix in quantum electrodynamics for a
charged (scalar) particle of mass $M$ and charge $Q$ in a uniform electric
field $\vec E$ is given by the proper time integral \cite{Loewe,AMSS}
\bea
\rho(\vec y,\vec y;\beta)&=&\frac\beta2\,\int\limits_0^\infty\frac{d\tau}
\tau~\frac{\e^{-M^2\tau/2}}{(2\pi\tau)^{\frac d2-1}}\,\frac{QE}{4\pi
\sin\frac{QE\tau}2}\label{rhopart}\\
&&\times\left.\Theta_3\left(\frac{i\beta Q\vec E
\cdot\vec y}{2\pi}\,\right|\frac{i\beta^2QE}{4\pi\tan\frac{QE\tau}2}
\right) \ . \nonumber
\eea
This result shows that the free energy of the system is trivial, in that the
integration of (\ref{rhopart}) over $\vec y$ picks up only the ground state of
winding number $n=0$ (due to the occurence of the theta-function $\Theta_3$).
Note that the density matrix (\ref{rhopart}) is complex-valued and its
imaginary part gives the Schwinger probability amplitude (\ref{Schwinger}).
Naively, the translational symmetry which is broken by the external electric
field could be restored by choosing the temporal gauge for the gauge potential
rather than the static gauge. However, this gauge choice ruins the global gauge
invariance of the system. The gauge potential in this case is not a function on
spacetime because it is multi-valued under periodic shifts around the
temperature circle, and it can only be properly defined with respect to a local
covering of the thermal direction. Requiring that the theory be independent of
the choice of covering requires the addition of generalized Wu-Yang terms to
the action (The mathematical details of this construction can be found in
\cite{AMSS}). These terms restore gauge invariance, but they also reinstate
precisely the same $\vec y$-dependent factor in (\ref{rhopart}). Therefore, the
gauge-invariant free energy remains trivial.

For a thermal state of charged strings, the constraint (\ref{zeromodeint})
forces us to take $\vec F=\vec0$. This selects the constant gauge field
configuration $A_0(x)=a_0$. Although this field is pure gauge, it can only be
removed by a singular gauge transformation. Therefore, charged states will
still depend on $a_0$, or equivalently the canonical, gauge invariant momentum
of the open strings depends on $a_0$. The gauge field background cannot be
removed because there is a non-trivial gauge invariant holonomy
$\e^{i(e_2-e_1)na_0}$ which arises from the boundary terms in the action
(\ref{Sann}) in the sector of temperature winding number $n$. This holonomy is
simply the Polyakov loop operator for the annular geometry. We can therefore
study the free energy for this constant gauge field configuration and compute
the effective action for charged strings in a generic, time-independent
background gauge field $a_0(\vec x_0)$. This yields the free energy that is
required to introduce a heavy charged particle into the system and thereby
gives information about confinement, which is the pertinent property of the
Hagedorn transition.

The action is now given by adding to the first line of (\ref{SannT}) the term
$i(e_2-e_1)na_0$. After summing over all $n\in\zed$, the appropriate
modification of the one-loop vacuum energy (\ref{annvacuum}) is given by
\bea
{\cal F}_0&=&\frac1{(8\pi^2\alpha')^{13}}\,\int\limits_0^\infty ds~
\eta(2is)^{-24}\non&&\times\,\Theta_3\left(\frac{(e_2-e_1)\beta}{2\pi}
\,a_0\left|\frac{i\beta^2}{32\pi^2\alpha's}\right.\right) \ ,
\label{F0}\eea
where we have made a modular transformation $t=1/s$ (mapping the one-loop open
string annulus diagram onto the tree-level closed string cylinder diagram) and
used the Poisson resummation formula
$\eta(-\frac1\tau)=\sqrt{-i\tau}\,\eta(\tau)$. The integral (\ref{F0}) can be
evaluated explicitly by expanding the Dedekind function using the formula
\be
\prod_{n=1}^\infty\Bigl(1-\e^{-nl}\Bigr)^{-(d-2)}=\sum_{N=0}^\infty d_N^{\rm b}
{}~\e^{-Nl} \ ,
\label{dNbos}\ee
where $d_N^{\rm b}$ is the degeneracy of bosonic string states at level $N$.
For the first two levels we have $d_0^{\rm b}=1$ and $d_1^{\rm b}=d-2$. By
expanding the theta-function $\Theta_3$ in an infinite series we can thereby
perform the integral (\ref{F0}) and arrive at
\bea
{\cal F}_0&=&\frac{2\beta}{(8\pi^2\alpha')^{14}}\,\sum_{N=0}^\infty
\frac{d_N^{\rm b}}{\sqrt N}~\sum_{n=1}^\infty
n\,K_1\left(n\beta\,\sqrt{\frac N{2\alpha'}}~\right)\non&&\times\,
\cos n\beta(e_1-e_2)a_0 \ ,
\label{F0Bessel}\eea
where $K_1(z)$ is the irregular modified Bessel function of order 1. The $N=0$
contribution to (\ref{F0Bessel}) of course diverges because of the tachyonic
instability. The next contribution is from the level $N=1$, corresponding to
the 26-dimensional Yang-Mills multiplet, which is well-defined. From the
asymptotic expansion $K_1(z)\sim\e^{-z}\sqrt{\pi/2z}$ for $|z|\to\infty$, it
follows that its contribution is suppressed in the low-temperature limit
$\beta\gg\sqrt{\alpha'}$ by terms of order $\e^{-\beta/\sqrt{\alpha'}}$.

Nevertheless, this calculation illustrates the general feature whose
instabilities are cured by computing the one-loop free energy of the
superstring gas \cite{AMSS}. Then the lowest $N=0$ level yields a finite
contribution in the low temperature limit which corresponds to the ten
dimensional Yang-Mills supermultiplet. By including the tree-level Born-Infeld
actions for the disc amplitudes of the charged string endpoints, we arrive at
the total (normalized) effective action for the gauge field $a_0(\vec x_0)$ up
to one-loop order in the form
\bea
\Gamma[a_0]&=&\frac1{(2\pi\alpha')^2g_s\left(e_1^2+e_2^2\right)}\non&&\times\,
\int d\vec x_0~
\left[\sum_{k=1,2}\sqrt{1+\Bigl(2\pi\alpha'e_k\nabla a_0(\vec x_0)\Bigr)^2}
\right.\non&&+\,\frac{\mu^2}{\beta^2(e_1-e_2)^2}\,\cos\beta(e_1-e_2)
a_0(\vec x_0)\non&&+\Biggl.{\cal O}\left(g_s^4\,,\,
\e^{-\beta/\sqrt{\alpha'}}~\right)\Biggr] \ ,
\label{Gammaa}\eea
where
\be
\mu^2=3\pi^2\cdot2^{22}g_s^2(\alpha')^3\beta^{-8}\left(e_1^2+e_2^2
\right)(e_1-e_2)^2 \ .
\label{Debyemass}\ee
Thus the low temperature modification of the Born-Infeld action is a
generalization of the sine-Gordon theory representation of the classical
Coulomb gas where the standard kinetic term for the gauge field is replaced by
the Born-Infeld Lagrangian. The main feature of this field theory is that the
linearized equation of motion for the minima of the free energy (\ref{Gammaa})
takes the form $(-\nabla^2+\mu^2)a_0(\vec x_0)=0$, which has exponentially
decaying solutions $a_0(\vec x_0)\sim\e^{-\mu|\vec x_0|}$ for $|\vec
x_0|\to\infty$. In this approximation the constant (\ref{Debyemass}) appears as
a mass term for the gauge field in (\ref{Gammaa}) and acts like a
Debye screening mass. It is clear for this reason that constant electric fields
cannot be extrema of the effective action, i.e. the existence of uniform
electric fields is inconsistent with the existence of a Debye mass. Note that
(\ref{Debyemass}) vanishes for neutral string states.

As expected, the Debye mass (\ref{Debyemass}) is the same as the one that would
arise in ordinary ten dimensional Yang-Mills theory. The contributions from
massive string states are exponentially suppressed by terms of order
$\e^{-\beta/\sqrt{\alpha'}}$. Since $\mu^2\propto T/T_{\rm H}(0)$, for
temperatures well below the critical Hagedorn temperature the Debye mass is
small and electric fields become more and more long-ranged. Stringy effects
essentially only play a role at temperatures near the Hagedorn transition.
Furthermore, the Born-Infeld generalization (\ref{Gammaa}) of the sine-Gordon
model has solitons which generalize the solitary waves of the plasma phase of
the ordinary Coulomb gas \cite{AMSS}. In gauge field theories these solitons
exist as $\zed_N$ domain walls. This characterization could prove useful for
other aspects of the Hagedorn transition in background fields.

\section{D-brane dynamics}

$T$-duality maps free open strings to open strings whose endpoints are attached
to D-branes. It replaces the quantities $\partial_ax^i$ by
$i\epsilon_{ab}\partial_bx^i$ and Neumann boundary conditions for the string
embedding fields $x^i$ with Dirichlet ones. The results of the previous
sections may be interpreted directly as the appropriate contributions to the
tension of a D9-brane with some background field distribution. $T$-dualizing
these expressions in $9-p$ of the spacetime directions is then tantamount in
string perturbation theory to adding an extra open string mass factor
$t^{-p/2}\,\e^{-r^2t/2\pi\alpha'}$ to the Teichm\"uller integration measure,
reflecting the Dirichlet nature of the $9-p$ transverse directions, where $r$
is the separation between parallel branes. If the background field is constant,
then its components which do not lie along the D$p$-brane can be gauged away.
The open string annulus amplitude then becomes the one-loop effective potential
between two D$p$-branes with generic background fields on each brane. Such a
configuration describes a boundary condensate of the stretched open strings
between the branes in the electromagnetic field. By keeping the transverse
electric field component non-vanishing, we may also describe the interaction
potential between moving branes. Much of the analysis we have made thus far for
the problem of open strings in electromagnetic fields has dual analogs for
D-brane dynamics. However, in the D-brane picture many of the stringy effects
that we have unveiled for the external field problem have very natural
dynamical explanations. In this section we will use the external field problem
to describe the dynamics of D-branes. We will restrict our attention to
D0-branes for simplicity.

Under $T$-duality, electric fields map onto the trajectories of D-branes as
follows. The coupling $\exp ie\oint\vec A(x^0)\cdot\partial_\parallel\vec x$ of
a time-varying, spatially constant electric field $\vec E=\partial_0\vec A$ to
a boundary that carries charge $e$ is replaced by the vertex operator
$\exp\frac1{2\pi\alpha'}\oint\vec y(x^0)\cdot\partial_\perp\vec x$ for a moving
D0-brane \cite{DLP,Leigh} travelling with velocity $\vec v=\partial_0\vec
y=2\pi\alpha'e\vec E$. The $\beta$-function equations for this coupling can be
interpreted as the classical equations of motion for the 0-brane. Constant
electric field thereby corresponds to uniform motion of the branes while a
neutral string would represent a pair of branes moving with zero relative
velocity. In string perturbation theory the electric field and moving D-brane
problems are identical because of the perturbative duality between Neumann and
Dirichlet boundary conditions on the string fields \cite{Leigh}. In the former
case the effective dynamics for a slowly varying electric field is governed by
the Born-Infeld action. Under $T$-duality this action simply maps onto the
usual action for a relativistic point particle,
\be
S_{\rm D0}=\int d\tau~T_0\,\sqrt{1-\vec v(\tau)^2} \ ,
\label{SD0}\ee
where $T_0=1/g_s\sqrt{\alpha'}$ is the 0-brane tension, i.e. the BPS mass of
the D-particles. The Born-Infeld action is the non-trivial result of a
resummation of all stringy order $\alpha'$ corrections, and among other things
it leads to a limiting value $E_c=(2\pi\alpha'e)^{-1}$ of the external electric
field, above which the system becomes unstable. In the dual picture, this is
simply a consequence of the laws of relativistic particle mechanics for the
0-brane, with the critical velocity corresponding to the speed of light. At
this velocity, we can make a large boost to bring the brane to rest, so that in
the $T$-dual picture of the original string theory this amounts to boosting to
large momentum. Thus the string theory with the electric background near the
critical limit is equivalent to the string theory in the infinite momentum
frame, or equivalently in the formalism of discrete light-cone quantization.

A planar static D-brane is a BPS defect that preserves half of the original
spacetime supersymmetries. D-branes in supersymmetric configurations exert no
static force on each other because supersymmetry ensures that the Casimir
energy of the stretched open strings vanishes \cite{Pol95}. Uniform motion of a
single D-brane cannot have any non-trivial consequences, because it depends on
the choice of an inertial frame. However, setting a pair of branes in relative
motion breaks the supersymmetry of the system and a velocity dependent
potential appears between them.

Another non-BPS configuration of D-branes is that which lives in a thermal
state of the theory. In the supergravity picture, non-extremal branes have a
natural temperature and entropy. The Hawking radiation of a certain class of
near extremal black holes with Ramond-Ramond charge may be interpreted in terms
of the emission of closed string modes by a thermal state of D-branes
\cite{SV,CM,MS}. It has been suggested \cite{BFKS} that the gravitational
Hawking temperature and the temperature of a Boltzmann gas of D0-branes should
be identified. This is based on the conjecture \cite{Maldacena} that the black
hole free energy resulting from classical supergravity is described accurately
by the large 't~Hooft coupling limit of the supersymmetric matrix quantum
mechanics describing the dynamics of $N$ D-particles \cite{BFSS}. In this
model, which comes from the leading Yang-Mills reduction of the (non-abelian)
Born-Infeld action, the brane coordinate fields are $N\times N$ Hermitian
matrices whose eigenvalues represent the collective coordinates of the
D-particles, while the off-diagonal fluctuations represent the Higgs fields
corresponding to short open string excitations between the parallel branes
\cite{Witten1}. The breaking of supersymmetry is then explicit in the fact that
the thermal partition function is computed with periodic temporal boundary
conditions  for the boson fields and anti-periodic ones for the fermion fields.
The model accurately describes the leading velocity corrections to the
tree-level action (\ref{SD0}) \cite{DKPS}, so it is natural to use it to
describe the thermodynamics of moving D-branes. In this section we will
describe some new calculations which compute these corrections to the static
D-brane amplitudes.

\subsection{Velocity dependent forces}

In this subsection we will present a novel derivation, using the Polyakov path
integral, of the known formula \cite{Bachas,BCDi} for the one-loop vacuum
energy of D-branes moving with uniform velocity. Path integral treatments of
D-branes can also be found in \cite{Kar1}. Consider a D-string in the
presence of a boundary condensate of constant electric field $E$ in Type IIB
superstring theory. The one-loop correction to the effective Born-Infeld action
at tree-level comes from the annulus string diagram which describes two
D1-branes with an open string stretching between them. Neumann boundary
conditions are taken along the axes 0,1, and Dirichlet ones along the
transverse axes $2,\dots, d-1$. These latter directions will be labelled
collectively in what follows by the superscript $\perp$. The open string
parametrization is $0\leq\sigma\leq1$, $0\leq\tau\leq t$, where $\ln t$ is the
Teichm\"uller parameter of the annulus. The string carries charges $e_\sigma$
at the worldsheet boundaries $\sigma=0,1$. The Euclidean action is analogous to
(\ref{Sstrip}),
\bea
S_{\rm D1}&=&S-ie_0E\left.\int\limits_0^t
d\tau~x^1\,\d_\tau x^0\right\vert^{\sigma=0}
\non& &+\left.ie_1E\int\limits_0^t
d\tau~x^1\,\d_\tau x^0 \right\vert^{\sigma=1} \ ,
\label{actionD1}\eea
where the bulk action $S$ is given by
\be
S=\frac 1{4\pi\alpha^\prime}\,\int\limits_0^td\tau~\int\limits_0^1d \sigma~
\left( \d_\tau x^\mu\,\d_\tau x_\mu
+\d_\sigma x^\mu\,\d_\sigma x_\mu \right) \ .
\label{bulkaction}\ee
Varying the action (\ref{actionD1}) leads to the boundary conditions
\bea
\left(\d_\sigma x^0 -2\pi i\alpha^\prime e_0E\,
\d_\tau x^1\right)_{\sigma=0}=0 \ , \non
\left(\d_\sigma x^0 -2\pi i\alpha^\prime e_1E\,
\d_\tau x^1\right)_{\sigma=1}=0 \ , \non
\left(\d_\sigma x^1 -2\pi i\alpha^\prime e_0E\,
\d_\tau x^0\right)_{\sigma=0}=0 \ , \non
\left(\d_\sigma x^1 -2\pi i\alpha^\prime e_1E\,
\d_\tau x^0\right)_{\sigma=1}=0 \ , \non
\left.\d_\tau x^\perp\right\vert_{\sigma=0} =0
{}~~\hbox{or}~~\left.x^\perp\right\vert_{\sigma=0}=y^\perp_0 \ , \non
\left.\d_\tau x^\perp\right\vert_{\sigma=1} =0
{}~~\hbox{or}~~\left.x^\perp\right\vert_{\sigma=1}=y^\perp_1 \ .
\label{e.b.c.}\eea

\subsubsection{Bosonic case}

If the coordinate $x^1\equiv x_{\rm N}^1$ is compactified on a circle of
circumference $L_{\rm N}$, then we can make a $T$-duality transformation along
the 1-axis which interchanges the Neumann and Dirichlet boundary conditions for
the open string. The new coordinate $x_{\rm D}^1$ takes values on a dual circle
of circumference $L_{\rm D}=4\pi^2\alpha'/L_{\rm N}$ with $\d_\tau x^1_{\rm N}
= \d_\sigma x^1_{\rm D}$ and $\d_\sigma x^1_{\rm N} = \d_\tau x^1_{\rm D}$. The
boundary conditions (\ref{e.b.c.}) can be rewritten as
\bea
\left.\d_\sigma \left(x^0 - \v_0  x^1\right)
\right\vert_{\sigma=0}=0 \ , \non
\left.\d_\sigma \left(x^0 - \v_1  x^1\right)
\right\vert_{\sigma=1}=0 \ , \non
\left.\d_\tau \left(x^1 + \v_0x^0 \right)
\right\vert_{\sigma=0}=0 \ , \non
\left.\d_\tau \left(x^1 + \v_1x^0 \right)
\right\vert_{\sigma=1}=0 \ , \non
\left.\d_\tau x^\perp\right\vert_{\sigma=0} =0
{}~~\hbox{or}~~\left.x^\perp\right\vert_{\sigma=0}=y^\perp_0 \ , \non
\left.\d_\tau x^\perp\right\vert_{\sigma=1} =0
{}~~\hbox{or}~~\left.x^\perp\right\vert_{\sigma=1}=y^\perp_1 \ ,
\label{Ev.b.c}\eea
where we have simply denoted $x^1\equiv x_{\rm D}^1$, and
\be
u_\sigma=iv_\sigma=2\pi i\alpha'e_\sigma E~~,~~\sigma=0,1
\label{velocity}\ee
are the analytic continuations of the velocities of the two string endpoints
from Minkowski to Euclidean space arising from Wick rotation of both $x^0$ and
$\tau$.

In the case of static D0-branes ($v_\sigma=0$), the mode expansions of the
string fields which diagonalize the action and solve the boundary conditions
are well-known to be given by~\cite{Min78}
\bea
x_{(0)}^0(\tau,\sigma)& =& \sum_{m=-\infty}^\infty~\sum_{n=0}^\infty
a_{mn}~\e^{2\pi i m \tau /t}\,\cos{\pi n \sigma} \ , \non
x_{(0)}^1(\tau,\sigma)& =& y^1_0+\left(y^1_1-y^1_0\right)\sigma\non& &+\,
\sum_{m=-\infty}^\infty~\sum_{n=1}^\infty
b_{mn}~\e^{2\pi i m \tau /t}\,\sin{\pi n \sigma} \ , \non
x_{(0)}^\perp(\tau,\sigma)& =& y^\perp_0+\left(y^\perp_1-y^\perp_0
\right)\sigma\label{smodev=0}\\& &+\,\sum_{m=-\infty}^\infty~\sum_{n=1}^\infty
x^\perp_{mn}~\e^{2\pi i m \tau /t}\,\sin{\pi n \sigma} \ , \nonumber
\eea
where $a_{-mn}=a_{mn}^*$, $b_{-mn}=b_{mn}^*$, and
$x^\perp_{-mn}=(x^\perp_{mn})^*$. The mode expansions which solve the boundary
conditions (\ref{Ev.b.c}) may then be obtained by rotating the fields
(\ref{smodev=0}) through angle $\pi\alpha_0+\pi\alpha\sigma$ in the 0--1 plane
to get
\bea
x^0(\tau,\sigma)& =&
\cos{(\pi \alpha_0 +\pi \alpha \sigma)}\,x_{(0)}^0(\tau,\sigma)\non& &
+\,\sin{(\pi \alpha_0 +\pi \alpha \sigma)}\,x_{(0)}^1(\tau,\sigma) \ , \non
x^1(\tau,\sigma)& =&-\sin{(\pi \alpha_0 +\pi \alpha \sigma)}\,
x_{(0)}^0(\tau,\sigma)\non& &+\,\cos{(\pi \alpha_0 +\pi \alpha \sigma)}
\,x_{(0)}^1(\tau,\sigma) \ , \non
x^\perp(\tau,\sigma)& =&x_{(0)}^\perp(\tau,\sigma) \ .
\label{smode}\eea
The fields (\ref{smode}) obey the boundary conditions~\rf{Ev.b.c} with the
identifications of the rotation angles as the rapidities of the Euclidean
boost,
\be
\pi\alpha_\sigma= \arctan\v_\sigma \ , \ \sigma=0,1 \ ,~~
\alpha=\alpha_1-\alpha_0 \ .
\label{alpha's}\ee
The mode expansion (\ref{smode}) diagonalizes the action (\ref{bulkaction})
which enters the Euclidean path integral
\bea
{\cal F}&=&2\,\frac{(2\alpha')^{-d/2}}\pi\non&&\times\,
\int\limits_0^\infty \frac{dt}t~\int Dx^\mu~D({\rm ghosts})~
\e^{-S-S_{\rm ghosts}}
\label{pathint}\eea
with the boundary conditions~\rf{Ev.b.c}. Here an extra factor of 2 has been
inserted to account for the symmetry under interchange of the two endpoints of
an oriented string~\cite{Pol95}.

Let us begin by evaluating the contributions from the non-zero modes of the
fields $x^0$ and $x^1$ to the path integral (\ref{pathint}). The modes with
indices $(m,n)$ and $(m',n')$ are orthogonal for $m'\neq-m$, and so their
contribution to the action (\ref{bulkaction}) is
\bea
S_{\rm q}&=&\frac{\pi t}{8\alpha^\prime}\,
\sum_{m=-\infty}^\infty~\sum_{n=1}^\infty
\left[\left(\alpha^2+n^2+\frac{4m^2}{t^2}\right)\right.\non& &\times\,
\biggl(a_{mn}a_{-mn}+b_{mn}b_{-mn}\biggr)\non& &
-\left.2\alpha n \biggl(a_{mn}b_{-mn}+a_{-mn}b_{mn}\biggr)\right] \ ,
\label{Sq}\eea
where we have used the fact that the modes with either sine or cosine functions
are orthogonal over the semi-period of $\sigma$, while the cross terms which
mix sine and cosine functions do not occur. Evaluating the resulting Gaussian
functional integral in (\ref{pathint}) over these non-zero modes produces the
determinant
\bea
&&\prod\limits_{m,n}
\left|\matrix{(\alpha^2+n^2)t+\frac{4m^2}t & -2\alpha n t \cr
-2\alpha n t& (\alpha^2+n^2)t+\frac{4m^2}t\cr}\right|^{-1/2} \non
&&~~=\prod\limits_{m=-\infty}^\infty~\prod\limits_{n=1}^\infty
\left((n+\alpha)\sqrt{t}+2im/\sqrt{t}\,\right)^{-1}\non&&~~~~~\times\,
\left((n-\alpha)\sqrt{t}+2im/\sqrt{t}\,\right)^{-1}\non
&&~~=\prod\limits_{n=1}^\infty\frac 1{4
\sinh{\frac{\pi(n+\alpha)t}2}\,
\sinh{\frac{\pi(n-\alpha)t}2}} \ ,
\label{bos}\eea
where we have used the formula
\be
\prod_{m=-\infty}^\infty\Bigl(my+x\Bigr)=\sin\frac{\pi x}y \ .
\label{sinprodformula}\ee
In arriving at (\ref{bos}) we have ignored an overall constant factor which may
be set equal to unity by using zeta-function regularization (\ref{zetafnreg})
of the infinite product.

Analogously, the contribution to~\rf{bulkaction} from the zero modes of the
fields $x^0$ and $x^1$ is
\bea
S_0&=&\frac{\pi t}{4\alpha^\prime}\,\sum\limits_{m=-\infty}^\infty
\left( \alpha^2+\frac{4m^2}{t^2} \right)a_{m0}\,a_{-m0}\label{zeroS}\\& &
-\,\frac{\alpha t}{2\alpha^\prime}\,a_{00}\left(y^1_1-y^1_0\right)
+ \frac t{4\pi\alpha^\prime}\,\left(y^1_1-y^1_0\right)^2 \ . \nonumber
\eea
The corresponding Gaussian functional integral gives
\be
\prod\limits_{m=-\infty}^\infty \left(
\alpha^2 t + 4m^2/t\right)^{-1/2} = \frac 1{2\sinh{\frac{\pi\alpha t}2}} \ .
\label{bos0}\ee
For any non-vanishing $\alpha$ the distance $y^1_1-y^1_0$ between the
ends of the string along the direction of motion can be absorbed into the
quantity $a_{00}$ and disappears from the final result, as it should. If
$\alpha=0$ the integral over $a_{00}$ produces the volume along the 0-axis and
the last
term on the right-hand side of \eq{zeroS} remains just as for the
transverse directions which contribute the usual quantity
\bea
&&\prod\limits_{n=1}^\infty
\frac 1{\left(2\sinh{\frac{\pi nt}2}\right)^{d-2}}
{}~\e^{-\frac t{4\pi\alpha^\prime}(y^\perp_1-y^\perp_0)^2  }\non
&&~~=\frac1{\eta\left(\frac{it}2 \right)^{d-2}}
{}~\e^{-\frac t{4\pi\alpha^\prime}(y^\perp_1-y^\perp_0)^2  } \ .
\label{bostransv}\eea

We now multiply the three quantities (\ref{bos}), (\ref{bos0}) and
(\ref{bostransv}) together, take into account the contributions from the
conformal ghost fields, and use the identity
\bea
&&\frac 1{2\sinh{\frac{\pi\alpha t}2}}\,
\prod\limits_{n=1}^\infty
\frac {\sinh^2{\frac{\pi nt}2}}
{\sinh{\frac{\pi(n+\alpha)t}2}\,
\sinh{\frac{\pi(n-\alpha)t}2}}\non
&&~~=\frac1{2\pi}\,\frac{\Theta^\prime_1\left(0\left\vert
\frac{it}2\right.\right)}{\Theta_1\left(\frac{i\alpha t}2\left\vert
\frac{it}2\right.\right)} \ .
\eea
In this way we find that the vacuum energy functional (\ref{pathint}) is given
by
\be
{\cal F}=\frac1{2\pi}\,\int\limits_0^\infty \frac{dt}t ~
\e^{-\frac t{4\pi\alpha^\prime}(y^\perp_1-y^\perp_0)^2  }\,
\frac{\Theta^\prime_1\left(0\left\vert
\frac{it}2\right.\right)}{\Theta_1\left(\frac{i\alpha t}2\left\vert
\frac{it}2\right.\right)}\,
\frac1{\eta\left(\frac{it}2 \right)^{24}}
\label{UUF}\ee
in the critical dimension $d=26$. This coincides with the result
of Refs.~\cite{Bachas,BCDi} for the bosonic string which is expressed in terms
of the Minkowski space rapidity $\epsilon=i\alpha$. Note that, in the present
derivation, we have simply used the action~\rf{bulkaction} without explicitly
adding boundary terms to correctly reproduce the bosonic amplitude for
moving D0-branes.

\subsubsection{RNS formulation}

We will now turn to the superstring vacuum amplitude. The open string is again
parametrized by the worldsheet coordinates $(\tau,\sigma)$. On the annulus
there are four standard spin structures, $(+,-)$, $(+,+)$ in the R-sector
(associated with spacetime fermions) and $(-,-)$, $(-,+)$ in the NS-sector
(associated with spacetime bosons), which represent the periodicities of the
worldsheet fermion fields with respect to $(\tau,\sigma)$. The GSO projection
dictates that the physical amplitude is obtained by summing over the
contributions from the four spin structures. The fermionic part of the
superstring action is given by
\bea
S_{\rm F}&=&\frac i2\,\int\limits_0^td\tau~\int\limits_0^1d\sigma~\Bigl(
\psi^\mu(\d_\tau+i\d_\sigma)\psi^\mu\Bigr.\non&&+\left.
\tilde\psi^\mu(\d_\tau-i\d_\sigma) \tilde\psi^\mu\right)
\label{RNSfermionic}\eea
where $\psi^\mu$ and $\tilde\psi^\mu$ are Grassmann-valued fields which
transform as $SO(8)$ vectors.

In the absence of an external field, or for static D-branes, the fermion fields
obey the standard superstring boundary conditions
\bea
\psi_{(0)}^\mu(\tau,0)&=&\tilde\psi_{(0)}^\mu(\tau,0) \ , \non
\psi_{(0)}^\mu(\tau,1)&=&(-1)^a\,\tilde\psi_{(0)}^\mu(\tau,1) \ ,
\label{RNSbc}\eea
where $a=0$ for the R-sector and $a=1$ for the NS-sector.
The mode expansions in the $(\pm,-)$ sectors are given by
\bea
\psi_{(0)}^\mu(\tau,\sigma)&=&\sum\limits_{m,n=-\infty}^\infty
\psi^\mu_{mn}~\e^{\pi i(2m+1)\tau/t}~\e^{\pi i(n+\frac a2)\sigma} \ ,
\non \tilde\psi_{(0)}^\mu(\tau,\sigma)&=&\sum\limits_{m,n=-\infty}^\infty
\psi^\mu_{mn}~\e^{\pi i(2m+1)\tau/t}~\e^{-\pi i(n+\frac a2)\sigma} \ ,
\non&&
\label{pm-modes}\eea
while in the $(\pm,+)$ sectors they are
\bea
\psi_{(0)}^\mu(\tau,\sigma)&=&\sum\limits_{m,n=-\infty}^\infty
\psi^\mu_{mn}~\e^{2\pi im\tau/t}~\e^{\pi i(n+\frac a2)\sigma} \ , \non
 \tilde\psi_{(0)}^\mu(\tau,\sigma)&=&\sum\limits_{m,n=-\infty}^\infty
\psi^\mu_{mn}~\e^{2\pi im\tau/t}~\e^{-\pi i(n+\frac a2)\sigma} \ .
\non&&
\label{pm+modes}\eea
For moving branes we rotate the fields $\psi_{(0)}^\mu$ with respect to
$\tilde\psi_{(0)}^\mu$ by the orthogonal matrix~\cite{GG}
\be
M_{\mu\nu}(\sigma)=\e^{(\pi \alpha_0+\pi \alpha \sigma) \Sigma^{01}_{\mu\nu}}
\label{vectororthogonal}\ee
where $\Sigma^{\rho\lambda}_{\mu\nu}=\delta^\rho_\mu\,\delta^{\lambda}_\nu-
\delta^\lambda_\mu\,\delta^{\rho}_\nu$ are the generators of $SO(8)$
rotations in the vector representation
and the rapidities $\alpha$ are given by \eq{alpha's}. Since
\bea
M_{\mu\nu}(0)&=&\e^{\pi \alpha_0 \Sigma^{01}_{\mu\nu}}=
\left(\frac{\id+\v_0\Sigma^{01}}{\id-\v_0\Sigma^{01}}\right)_{\mu\nu} \ , \non
M_{\mu\nu}(1)&=&\e^{\pi \alpha_1 \Sigma^{01}_{\mu\nu}}=
\left(\frac{\id+\v_1\Sigma^{01}}{\id-\v_1\Sigma^{01}}\right)_{\mu\nu} \ ,
\eea
the rotation of the fermion fields by~\rf{vectororthogonal} solves the boundary
conditions
\bea
&&\psi_\mu(\tau,0)-\tilde\psi_\mu(\tau,0)\non&&~~=
\v_0\Sigma^{01}_{\mu\nu}\left(\psi^\nu(\tau,0)+
\tilde\psi^\nu(\tau,0)\right) \ , \non&&
\psi_\mu(\tau,1)-(-1)^a\,\tilde\psi_\mu(\tau,1)\non&&~~=
\v_1\Sigma^{01}_{\mu\nu}\left(\psi^\nu(\tau,1)+
(-1)^a\,\tilde\psi^\nu(\tau,1)\right) \ ,
\eea
which extend~\rf{RNSbc} to finite velocity
provided that~\rf{alpha's} is satisfied. The mode expansions are thereby given
from (\ref{pm-modes},\ref{pm+modes}) as
\bea
\psi_\mu(\tau,\sigma)&=&M_{\mu\nu}(\sigma)^{1/2}\,\psi_{(0)}^\nu(\tau,\sigma)
 \ , \non\tilde\psi_\mu(\tau,\sigma)&=&M_{\mu\nu}(\sigma)^{-1/2}\,
\tilde\psi_{(0)}^\nu(\tau,\sigma) \ ,
\label{psimodes}\eea
where here only the 0 and 1 components of the fields are rotated since the
boost is in the 0--1 plane. The mode expansions (\ref{psimodes}) diagonalize
the action~\rf{RNSfermionic} in each of the four worldsheet sectors. We shall
analyse the contributions first from each sector separately.

\noindent
$\underline{(+,-)~{\rm sector:}}$ Using (\ref{pm-modes}) with $a=0$ and
substituting (\ref{psimodes}) into (\ref{RNSfermionic}) leads to the action
\bea
S_{\rm F}&=&-\pi\sum\limits_{m=-\infty}^\infty~
\sum\limits_{n=-\infty}^\infty\psi^\mu_{-m-n}\non&&\times\,
\left( 2m+1 +t (in+\alpha\Sigma^{01})\right)_{\mu\nu}\psi^\nu_{mn} \ .
\eea
The functional Gaussian integral over the Grassmann variables $\psi_{mn}^\mu$
thereby produces the determinant
\bea
&&\prod\limits_{m,n=-\infty}^\infty
\det\pmatrix{2m+1 +i n t & \alpha t\cr-\alpha t & 2m+1 +int\cr}^{1/2}
\non&&~~~~~\times\, \Bigl( 2m+1 +i nt \Bigr)^4
\nonumber \\ &&~~=\prod\limits_{m=-\infty}^\infty~
\prod\limits_{n=-\infty}^\infty
\Bigl( 2m+1 +it (n+\alpha) \Bigr)^{1/2}\non&&~~~~~\times\,
\Bigl( 2m+1 +it (n-\alpha) \Bigr)^{1/2}\,
\Bigl( 2m+1 +int\Bigr)^4\nonumber \\&&~~ =
2^5 \cosh \frac{\pi t\alpha}2\,\prod\limits_{n=1}^\infty
4 \cosh \frac{\pi t(n+\alpha)}2\non&&~~~~~\times\,\cosh \frac{\pi t(n-\alpha)}2
\,\left(2 \cosh \frac{\pi tn}2 \right)^8 \non && ~~=
\frac{\Theta_2\left(\frac{i\alpha t}2\left\vert\frac{it}2\right.\right)}
{\eta\left(\frac{it}2\right)}\,\left(
\frac{\Theta_2\left(0\left\vert\frac{it}2\right.\right)}
{\eta\left(\frac{it}2\right)} \right)^4 \ ,
\eea
where we have used the formula
\be
\cos\pi x=\prod_{k=0}^\infty\left[1-\frac{4x^2}{(2k+1)^2}\right] \ .
\label{cosprod}\ee

\noindent
$\underline{(+,+)~{\rm sector:}}$ For $a=0$ the fermion fields (\ref{pm+modes})
have a zero mode, and so the contribution from this sector vanishes as usual
due to the integration over the fermionic zero mode.

\noindent
$\underline{(-,-)~{\rm sector:}}$ Setting $a=1$ in (\ref{pm-modes}) and
substituting (\ref{psimodes}) into (\ref{RNSfermionic}) leads to the action
\bea
S_{\rm F}&=& -\pi\sum\limits_{m=-\infty}^\infty~
\sum\limits_{n=-\infty}^\infty\psi^\mu_{-m-n}\\&&\times\,
\left[2m+1 +t\left(i(n+\mbox{$\frac12$})+\alpha\Sigma^{01}\right)
\right]_{\mu\nu}\psi^\nu_{mn} \ . \nonumber
\eea
The functional Gaussian integral over the Grassmann variables $\psi^\mu_{mn}$
gives
\bea
&&\prod\limits_{m=-\infty}^\infty~\prod\limits_{n=-\infty}^\infty
\Bigl( 2m+1 +it (n+\mbox{$\frac12$}+\alpha)\Bigr)^{1/2}\non && ~~~~~\times\,
\Bigl( 2m+1 +it (n+\mbox{$\frac12$}-\alpha)\Bigr)^{1/2}\non && ~~~~~\times\,
\Bigl( 2m+1 +it (n+\mbox{$\frac12$})\Bigr)^4
\nonumber \\&&~~ =  \prod\limits_{n=1}^\infty
4 \cosh \frac{\pi t(n-\frac12+\alpha)}2\,\cosh \frac{\pi t(n-\frac12-\alpha)}2
\non&&~~~~~\times\,\left(2 \cosh \frac{\pi t(n-\frac12)}2 \right)^8
\non && ~~=
\frac{\Theta_3\left(\frac{i\alpha t}2\left\vert\frac{it}2\right.\right)}
{\eta\left(\frac{it}2\right)}\,\left(
\frac{\Theta_3\left(0\left\vert\frac{it}2\right.\right)}
{\eta\left(\frac{it}2\right)} \right)^4 \ .
\eea

\noindent
$\underline{(-,+)~{\rm sector:}}$ Finally, by setting $a=1$ in (\ref{pm+modes})
the action reads
\bea
S_{\rm F}&=&-\pi\sum\limits_{m=-\infty}^\infty~
\sum\limits_{n=-\infty}^\infty\psi^\mu_{-m-n}\\&&\times\,
\left[2m +t\left(i(n+\mbox{$\frac12$})+\alpha\Sigma^{01}
\right)\right]_{\mu\nu}\psi^\nu_{mn} \ . \nonumber
\eea
The functional Gaussian integral over the Grassmann variables $\psi^\mu_{mn}$
yields
\bea
&&\prod\limits_{m=-\infty}^\infty~\prod\limits_{n=-\infty}^\infty
\Bigl( 2m +it (n+\mbox{$\frac12$}+\alpha) \Bigr)^{1/2}\non &&~~~~~\times\,
\Bigl( 2m +it (n+\mbox{$\frac12$}-\alpha) \Bigr)^{1/2}\,
\Bigl( 2m +it (n+\mbox{$\frac12$} )\Bigr)^4
\nonumber \\&&~~ =  \prod\limits_{n=1}^\infty
4 \sinh \frac{\pi t(n-\frac12+\alpha)}2\,\sinh \frac{\pi t(n-\frac12-\alpha)}2
\non&&~~~~~\times\,\left(2 \sinh \frac{\pi t(n-\frac12)}2 \right)^8
\non && ~~=
\frac{\Theta_4\left(\frac{i\alpha t}2\left\vert\frac{it}2\right.\right)}
{\eta\left(\frac{it}2\right)}\,\left(
\frac{\Theta_4\left(0\left\vert\frac{it}2\right.\right)}
{\eta\left(\frac{it}2\right)} \right)^4.
\eea

We now take into account the contributions from the conformal anti-ghost fields
in each of the three non-vanishing sectors above, sum over the spin structures
with weight $\frac12$, and multiply by the bosonic amplitude in the superstring
critical dimension $d=10$. In this way we arrive at the known formula
\cite{Bachas} for the superstring vacuum energy functional,
\bea
{\cal F}&=&\frac1{2\pi}\,\int\limits_0^\infty \frac{dt}t ~
\e^{-\frac t{4\pi\alpha^\prime}(y^\perp_1-y^\perp_0)^2  }\,
\frac{\Theta^\prime_1\left(0\left\vert
\frac{it}2\right.\right)}{\Theta_1\left(\frac{i\alpha t}2\left\vert
\frac{it}2\right.\right)}\,
\frac1{\eta\left(\frac{it}2 \right)^{12}}\non&& \times\,\frac 12\,\left[
\Theta_3\left(\frac{i\alpha t}2\left\vert\frac{it}2\right.\right)
\Theta_3\left(0\left\vert\frac{it}2\right.\right)^3 \right.\non &&
-\,\Theta_4\left(\frac{i\alpha t}2\left\vert\frac{it}2\right.\right)
\Theta_4\left(0\left\vert\frac{it}2\right.\right)^3\non&&
-\left.\Theta_2\left(\frac{i\alpha t}2\left\vert\frac{it}2\right.\right)
\Theta_2\left(0\left\vert\frac{it}2\right.\right)^3\right] \ .
\label{RNSUUF}\eea
By using the formula
\bea
&&\frac 12\,\Bigl[\Theta_3\left(\nu\left\vert{it}\right.\right)
\Theta_3\left(0\left\vert{it}\right.\right)^3
-\Theta_4\left(\nu\left\vert it\right.\right)
\Theta_4\left(0\left\vert{it}\right.\right)^3\Bigr.\non&&~~~~~
-\Bigl.\Theta_2\left(\nu\left\vert{it}\right.\right)
\Theta_2\left(0\left\vert{it}\right.\right)^3\Bigr]
=\Theta_1\left(\left.\frac \nu2 \right\vert it\right)^4
\eea
which is a consequence of the Riemann identity for Jacobi theta-functions, we
arrive at the result of Refs.~\cite{GG,BCDi} after a modular transformation
$t\mapsto1/t$ of the annular Teichm\"uller parameter.

\subsection{D-brane scattering}

After analytical continuation to Minkowski space, the quantity (\ref{RNSUUF})
can be interpreted as the forward scattering amplitude for two D-particles
moving with relative velocity $v=\tanh\pi\epsilon$ and impact parameter
$b=|y_1^\perp-y_0^\perp|$. It is a semi-classical result, in that the D-branes
are treated as classical sources and higher worldsheet topologies are
neglected, i.e. both the Compton wavelength and the Schwarzschild radius of the
D-branes are taken to vanish. The branes interact via virtual pairs of open
oriented strings which are stretched by the relative motion. The integrand of
(\ref{RNSUUF}) has an infinite number of poles along the real $t$-axis at
$t=2\pi(2n+1)/\epsilon$, where $n$ is an integer. These poles arise from the
zeroes of the trigonometric sine function in the product representation of the
theta-function $\Theta_1$. As a consequence, the vacuum energy acquires an
imaginary part which is given by the sum over the residues of the poles and
which gives the probability that the virtual strings materialize. This
phenomenon is simply the dual counterpart of the open string pair production in
a uniform background electric field that we described in section II.C. In the
present case the result has a much simpler interpretation. As the two
D-particles move away from each other, they continuously transfer their energy
to any open strings that stretch between them. A virtual pair of open strings
can therefore nucleate out of the vacuum and slow down or even completely stop
the relative motion.

The real part of the scattering amplitude also reveals a striking feature of
the low velocity dynamics of D-particles. The theta-functions
$\Theta_a(\nu|\tau)$ are even functions of $\nu$, and in the low velocity limit
$\Theta_1(\frac{\epsilon t}2|\frac{it}2)^4\sim{\cal O}(v^4)$. The absence of a
constant term in the velocity expansion of (\ref{RNSUUF}) is due to the
cancellation of the gravitational attraction and the Ramond-Ramond repulsion
for static D-branes \cite{Pol95}. However, not only the static, but also the
order $v^2$ force between two D-particles vanishes, i.e. identical Type II
D-branes do not scatter at non-relativistic velocities. Generally, the order
$v^2$ scattering of heavy solitons can be described by geodesic motion in the
moduli space of zero modes. Therefore, the moduli space of a pair of
D-particles, which at tree level is the flat quotient space
$\real^9\times\real^9/\zed_2$, remains completely flat to all orders in the
$\alpha'$ expansion. This is the dual statement of the fact that for maximally
supersymmetric gauge theories the Maxwell $F^2$ term in the effective action is
not renormalized. The next contribution comes at order $v^4$ in the velocity
expansion. The expansion of the effective action for two D-particles in their
velocities divided by powers of their separation $r$ is thereby given as
\bea
S_{\rm D0}&=&\int d\tau~\left[\frac{T_0}2\,\Bigl((\dot y_0^\perp)^2+
(\dot y_1^\perp)^2\Bigr)\right.\non&&-\left.
\frac{15}{16}\,(\alpha')^3\,\frac{v^4}{r^7}+\dots\right] \ .
\label{SD0vel}\eea
The $v^4$ potential in (\ref{SD0vel}) is the standard interaction term for
D0-branes in ten dimensional supergravity \cite{DKPS}.

The vacuum energy functional (\ref{RNSUUF}) could actually have been determined
using standard formulas for the partition functions of free massless fields
with twisted boundary conditions. The spectrum of an open string stretched
between two moving D-branes can be determined from the operatorial mode
expansions of the light-cone fields $x^\pm=\frac1{\sqrt2}(x^0\mp x^1)$ which
are given by
\bea
x^\pm(\tau,\sigma)&=&-i\,\sqrt{\frac{\alpha'}2}\,\sqrt{\frac{1\pm v_0}
{1\mp v_0}}\non&&\times\,\sum_{n=-\infty}^\infty\left(\frac{a_n^\pm}{n\pm
i\epsilon}~\e^{-\pi i(n\pm i\epsilon)(\tau+\sigma)}\right.\non&&+\left.
\frac{a_n^\mp}{n\mp
i\epsilon}~\e^{-\pi i(n\mp i\epsilon)(\tau-\sigma)}\right) \ .
\label{Xpmmode}\eea
It is easy to verify that $x^0$ and $x^1$ then obey the boundary conditions
(\ref{Ev.b.c}). In particular,
$x^\pm(\tau,1)=\e^{\pm\pi\epsilon}\,x^\pm(\tau,0)$, so that the two string
endpoints have relative velocity $v=\tanh\pi\epsilon$. Reality requires
$(a_n^\pm)^\dagger=a_{-n}^\pm$, while canonical quantization implies the
commutation relations $[a_n^+,a_m^-]=(n+i\epsilon)\,\delta_{n+m,0}$. The
D-brane motion modifies the vacuum energy as can be read off from the
light-like component of the total worldsheet Hamiltonian
\bea
L_0^\parallel&=&\frac{b^2}{4\pi^2\alpha'}+\sum_{n=1}^\infty(a_n^+)^\dagger
a_n^-+\sum_{n=0}^\infty(a_n^-)^\dagger a_n^+\non&&+\,
\frac{i\epsilon(1-i\epsilon)}2 \ ,
\label{L0light}\eea
where we have included the dependence on the impact parameter. Analogous mode
expansions arise for the worldsheet fermion fields. The relevant effect of the
brane motion on the stretched strings is to shift their oscillation frequencies
by $\pm i\epsilon$ in the boost plane and their energy by an overall velocity
dependent term. Similar expansions arise in the twisted sectors of orbifold
conformal field theories, with $i\epsilon$ identified as a real-valued rotation
angle. Therefore, in the operator formalism, the problem of moving D-branes is
formally identical to that of the stretched strings between the branes
belonging to a twisted sector of an orbifold string theory with imaginary twist
angle corresponding to the rapidity of the boost.

All of this is again completely analogous to the spectrum of free open strings
in a uniform electric field background, except for some important changes. The
expression (\ref{alpha's}) for the twist parameter $\epsilon=i\alpha$ has no
obvious interpretation in the electric field case, while here it is recognized
as the relativistic sum of the two brane velocities. Consistent with Lorentz
invariance, the spectrum only depends on the velocity $v$ of one brane in the
rest frame of the other. Furthermore, zero modes are omitted in the light-cone
mode expansions (\ref{Xpmmode}) to account for the fact that D-branes interact
locally in transverse space and in time. Keeping this and the orbifold analogy
in mind, it is straightforward to arrive at the annulus amplitude
(\ref{RNSUUF}). The orbifold interpretation further enables a very simple
calculation of the velocity dependent potential in (\ref{SD0vel}). In the
quasi-static approximation, which is valid to leading order in the inverse
separation, the potential is given simply as the sum of the ground state
energies of the corresponding harmonic oscillators as in (\ref{L0light}). There
are ten complex bosonic oscillators of frequencies $\omega_{\rm
b}^\perp=\sqrt{r^2}$ with multiplicity eight and $\omega_{\rm
b}^\pm=\sqrt{r^2\pm2iv}$ with multiplicity one each, where
$r=\sqrt{b^2+v^2\tau^2}$ is the distance between the branes at time $\tau$.
There are also two ghost oscillators each of frequency $\omega_{\rm
ghosts}=\sqrt{r^2}$, and 16 fermionic oscillators of frequencies $\omega_{\rm
f}^\pm=\sqrt{r^2\pm iv}$ with multiplicity eight each. The velocity dependent
potential is then given by
\bea
V(r)&=&8\,\omega_{\rm b}^\perp+\omega_{\rm b}^++\omega_{\rm b}^-\non&&-\,2\,
\omega_{\rm ghosts}-4\,\omega_{\rm f}^+-4\,\omega_{\rm f}^- \ .
\label{velpot}\eea
For $v=0$ the frequencies cancel and the static potential vanishes. For
$v\neq0$ we can expand each frequency as a power series in $r^{-1}$. At the
first three orders in $v/r^2$ the potential vanishes, while at fourth order the
energy between the 0-branes gives the expected leading result
$V(r)=-15v^4/16r^7+\dots$.

\subsection{Thermodynamics}

Just as we did with the thermal configuration of free open strings in
background electric fields, it is possible to demonstrate that there are no
excited states of a pair of moving D-branes with $v\neq0$ at finite temperature
\cite{AMSS}. Again the partition function picks up only the temperature
independent piece, and we may conclude that D-brane dynamics forbid uniform
velocity motion at finite temperature. The triviality comes from the same zero
mode operators, associated with the presence of a Wu-Yang term, as in the
electric field problem. Using $T$-duality, we may therefore attribute this
property of D-brane dynamics to the Debye screening of electric fields that we
discussed in section III.B.2. Just as Debye screening forbids constant electric
fields in open superstring theory, it also forbids the uniform motion of
D-branes. This implies that there is a damping of their motion analogous to
Debye screening. Recall that, in the dual picture, this is not the case for
constant {\it magnetic} fields. At one-loop order this corresponds to a
relative disalignment between a pair of branes, which is an allowed
configuration at finite temperature.

To investigate further the properties of D-brane dynamics at finite
temperature, one must consider appropriate non-uniform motion. This is a
difficult problem to treat using the usual, direct methods of string
perturbation theory, as is the dual problem for time-dependent background
fields. However, one can compute the thermodynamic free energy for moving
D0-branes by using the low-energy effective Yang-Mills theory description of
the D-brane dynamics \cite{Witten1,BFSS}. Then, a perturbative calculation will
be valid in the domain where $g_s^{1/3}\sqrt{\alpha'}\ll r$. We can therefore
effectively describe the thermodynamics in the limit of weak string coupling,
or equivalently when the branes are well separated. Since the D0-branes have
mass $T_0=1/g_s\sqrt{\alpha'}$ and are therefore very heavy in this limit, this
calculation will take into account the thermal fluctuations of the stretched
superstrings but not of the D-particles themselves.

The action is obtained from the dimensional reduction of ten-dimensional
maximally supersymmetric Yang-Mills theory to one temporal and zero spatial
dimensions,
\be
S_{\rm YM}[A,\Psi]=\frac1{g_{\rm YM}^2}\,\int d\tau~\tr\left(\frac14\,
F_{\mu\nu}^2+\frac i2\,\Psi\gamma^\mu D_\mu\Psi\right)
\label{SYM}\ee
where the Yang-Mills coupling constant $g_{\rm YM}$ is related to the string
coupling $g_s$ by $g_{\rm YM}^2=g_s/4\pi^2(\alpha')^{3/2}$. The gauge fields
$A_\mu(\tau)$ and the Majorana spinor fields $\Psi(\tau)$ depend only on the
time coordinate $\tau$. The diagonal components
\be
\vec y^{\,b}=2\pi\alpha'\vec a^{\,b}\equiv2\pi\alpha'\vec A^{\,bb}
\label{ybdef}\ee
of the gauge fields are interpreted as the position of the $b$-th D0-brane and
are treated as collective variables. The thermal partition function defines the
statistical mechanics of the gas of D0-branes through the path integral
\be
Z_{\rm YM}=\int D\vec y^{\,a}~\e^{-S_{\rm eff}[\vec y^{\,a}]}
\label{ZYM}\ee
with the Euclidean time coordinate $\tau$ compactified on a circle of
circumference $\beta=1/k_{\rm B}T$. The effective action for the D-particle
coordinates is constructed by integrating out the off-diagonal components of
the gauge fields, the fermion fields, and the Faddeev-Popov ghost fields
required for gauge fixing,
\bea
S_{\rm eff}[\vec y^{\,a}]&=&-\ln\int\prod_{a\neq b}Da_0^b~DA_\mu^{ab}
\label{SeffD0}\\&&\times\,\int D\Psi~
D({\rm ghosts})~\e^{-S_{\rm YM}-S_{\rm ghosts}} \ , \nonumber
\eea
with periodic boundary conditions for the gauge and ghost fields and
anti-periodic ones for the adjoint fermion fields around the compact
temperature direction. We will consider again only the case of a single pair of
D0-branes whose worldlines lie along the periodic temporal direction.

The integration in (\ref{SeffD0}) can be done in a simultaneous loop expansion
in the gauge theory and in a velocity expansion in the brane configurations.
This will produce meaningful results in the limit where $r\equiv|\vec
y^{\,1}-\vec y^{\,2}|$ is large and where $\vec v^{\,a}=\dot{\vec y}^{\,a}$ are
small (Dot denotes differentiation with respect to $\tau$). The one-loop
contribution can be obtained by expanding the action (\ref{SYM}) to second
order in the off-diagonal components of the gauge fields, and in the ghost and
fermion fields. The result of this standard Gaussian functional integration
produces a ratio of determinants of the form \cite{AMS}
\bea
S_{\rm eff}&=&\int\limits_0^\beta d\tau~\frac1{4g_{\rm YM}^2}\,\left(
(\dot{\vec a}^1)^2+(\dot{\vec a}^2)^2\right)\non&&+\,\frac12\,\tr_B\ln\Bigl[
\delta_{\mu\nu}\,{\cal D}^2+2i(f_{\mu\nu}^1-f_{\mu\nu}^2)\Bigr]
\non&&+\,\frac12\,\tr_B\ln\Bigl[\delta_{\mu\nu}\,{\cal D}^2-
2i(f_{\mu\nu}^1-f_{\mu\nu}^2)\Bigr]\non&&-\,2\,\tr_B\ln{\cal D}^2
-8\,\tr_F\ln{\cal D}^2 \ ,
\label{Seffdets}\eea
where we have introduced the second order differential operator on the
temperature circle,
\be
{\cal D}^2=(-i\partial_\tau+a_0)^2+|\vec a|^2 \ ,
\label{calDdef}\ee
which arises from the gauge covariant derivatives. Here
$a_\mu=a_\mu^1-a_\mu^2$, and we have included the tree-level term which gives
the non-relativistic kinetic energies of the D0-branes. The subscript $B$
(resp. $F$) indicates that the determinant is to be evaluated with periodic
(resp. anti-periodic) boundary conditions corresponding to the contributions
from the gauge and ghost (resp. adjoint fermion) fields, respectively. The
abelian field strength tensors $f_{\mu\nu}^b$ have non-vanishing components
$f_{0i}^b=\dot a_i^b$. The temporal components $a_0^b$ of the gauge fields may
be taken to be independent of the compactified time variable via the residual
abelian gauge invariance of the problem, and by periodicity to lie in the
interval $(-\frac\pi\beta,\frac\pi\beta]$. The determinants in (\ref{Seffdets})
have been evaluated for static D-brane configurations in \cite{AMS}. In what
follows we shall extend this computation to leading orders in the velocity
expansion for moving D-branes.

By using the proper time representation (\ref{lncalDint}) we are led to first
evaluate the trace
\be
\tr~\e^{-t{\cal D}^2}=\int\limits_0^\beta d\tau~\lim_{\tau'\to\tau}\,
\e^{-t{\cal D}^2}\,\delta(\tau-\tau') \ .
\label{trexpint}\ee
We will begin by computing the determinant in (\ref{Seffdets}) with bosonic
boundary conditions on the temporal circle. For this, we insert the periodic
delta-function
\be
\delta_B(\tau-\tau')=\frac1\beta\,\sum_{n=-\infty}^\infty\e^{-2\pi i
n(\tau-\tau')/\beta}
\label{deltaB}\ee
into (\ref{trexpint}), which incorporates the proper Matsubara frequencies and
gives
\be
\tr_B\,\e^{-t{\cal D}^2}=\frac1\beta\,\sum_{n=-\infty}^\infty~\int
\limits_0^\beta d\tau~\e^{-t({\cal A}_n+{\cal B}_n)}\cdot1 \ ,
\label{trBMatsubara}\ee
where we have introduced the operators
\bea
{\cal A}_n&=&\left(\frac{2\pi n}\beta+a_0\right)^2+|\vec a|^2 \ ,\non
{\cal B}_n&=&-i\partial_\tau
\left(-i\partial_\tau+\frac{2\pi n}\beta+a_0\right) \ .
\label{calAcalBdef}\eea
The expression (\ref{trBMatsubara}) is viewed as operating on a constant 1, and
the derivatives only contribute when they encounter terms involving $|\vec
a|^2$. Note that generally the position variables $\vec a^{\,b}(\tau)$ are only
periodic up to a permutation of the identical D-particles, which ensures that
the configuration of the coordinates is periodic. In the present case this
means that the relative coordinate $\vec a(\tau)$ can be either periodic or
anti-periodic. In both of these sectors, the distance $|\vec a(\tau)|$ is a
periodic function, and hence so is the operator ${\cal A}_n$.

To unravel the expression (\ref{trBMatsubara}), we use the generalization of
the Baker-Campbell-Hausdorff formula
\be
\e^{-t({\cal A}_n+{\cal B}_n)}=\e^{-t{\cal A}_n}~\e^{{\cal C}_n}~
\e^{-t{\cal B}_n}
\label{BCH}\ee
where
\bea
{\cal C}_n&=&-\frac{t^2}{2!}\,[{\cal A}_n,{\cal B}_n]\label{BCHcalC}\\&&-\,
\frac{t^3}{3!}\,\Bigl([{\cal A}_n,[{\cal A}_n,{\cal B}_n]]+
[[{\cal A}_n,{\cal B}_n],{\cal B}_n]\Bigr)+\dots \ . \nonumber
\eea
In the loop expansion we expand around the tree-level configuration whose
equation of motion is $\ddot{\vec a}=0$. The only non-vanishing time
derivatives of the operator ${\cal A}_n$ are then $\dot{\cal A}_n=2\vec
a\cdot\dot{\vec a}$ and $\ddot{\cal A}_n=2|\dot{\vec a}|^2$. Moreover, since
the integrand of (\ref{trBMatsubara}) is a periodic function, we can freely
integrate by parts and drop surface terms. Using (\ref{BCH},\ref{BCHcalC}) and
the equations of motion, we may compute the functional determinants in
(\ref{Seffdets}) to second order in the expansion in time-derivatives of $|\vec
a|$. After some tedious algebra, we arrive finally at
\bea
&&\tr_B\ln{\cal D}^2=\int\limits_0^\beta d\tau~\left[\frac1\beta\,
\ln\frac12\,\Bigl(\cosh\beta|\vec a|-\cos\beta a_0\Bigr)\right.\non&&+\,
\frac{|\dot{\vec a}|^2}{96|\vec a|}\,\frac{\sinh\beta|\vec a|}{\cosh\beta
|\vec a|-\cos\beta a_0}\label{trBlnDfinal}\\
&&\times\,\left\{\frac7{|\vec a|^2}-\,
\frac{7\beta}{|\vec a|\sinh\beta|\vec a|}\,\frac{1-\cosh\beta|\vec a|
\cos\beta a_0}{\cosh\beta|\vec a|-\cos\beta a_0}\right.\non&&+\left.\left.
\beta^2\,\frac{\cos\beta a_0(\cosh\beta|\vec a|+\cos\beta a_0)-2}
{(\cosh\beta|\vec a|-\cos\beta a_0)^2}\right\}+\dots\right] \ . \nonumber
\eea
where the ellipses denote terms which are of third and higher orders in
time-derivatives of ${\cal A}_n$. In arriving at (\ref{trBlnDfinal}) we have
evaluated the sum $\sum_n\ln{\cal A}_n$ using the formula
(\ref{sinprodformula}),\footnote{\baselineskip=12pt The equality
(\ref{sinprodformula}) is valid up to a function of $x$ which is unity at all
of the zeroes $x=-my$. This would produce different asymptotic behaviour in the
complex plane. Here we have defined it so that it is odd under reflection of
$x$ but not periodic under the shift $x\mapsto x+y$. There is no way to
preserve both of these symmetries, and this gives a simple example of an
anomaly.} and the sums $\sum_n{\cal A}_n^{-k}$ may then be computed from
the formula $(k-1)!\,{\cal A}_n^{-k}=-(-\partial/\partial|\vec a|^2)^k\ln{\cal
A}_n$. There are eight contributions of the form (\ref{trBlnDfinal}), one for
each of the directions transverse to the plane of motion. From this result we
must also subtract the fermionic contribution which comes from evaluating the
determinant with anti-periodic boundary conditions. The net effect of inserting
the anti-periodic delta-function into (\ref{trexpint}) is to replace $a_0$ by
$a_0+\frac\pi\beta$ everywhere, i.e. $\cos\beta a_0\to-\cos\beta a_0$ in
(\ref{trBlnDfinal}).

The final quantity we need is the velocity corrected determinant (the first two
determinants of (\ref{Seffdets})). This can be straightforwardly evaluated by
using the description given in the previous subsection of how the oscillator
frequencies are modified in the velocity-dependent potential between two
D0-branes. The leading order term $8\ln\frac{\cosh\beta|\vec a|-\cos\beta
a_0}{\cosh\beta|\vec a|+\cos\beta a_0}$ with no time derivatives is corrected
at finite velocity to
\bea
&&6\ln(\cosh\beta|\vec a|-\cos\beta a_0)\non&&~~
+\,\ln\left(\cosh\beta\sqrt{|\vec a|^2+2i|\dot{\vec a}|}-\cos\beta a_0
\right)\non&&~~+\,\ln\left(\cosh\beta\sqrt{|\vec a|^2-2i|\dot{\vec a}|}-
\cos\beta a_0\right)\non&&~~-\,4\ln\left(\cosh\beta\sqrt{|\vec a|^2+i|
\dot{\vec a}|}+\cos\beta a_0\right)\non&&~~-4\,\ln\left(\cosh\beta
\sqrt{|\vec a|^2-i|\dot{\vec a}|}+\cos\beta a_0\right) \ .
\label{leadingcorrect}\eea
By summing the expansion of (\ref{leadingcorrect}) to second order in the
velocity $|\dot{\vec a}|$ and the eight order $|\dot{\vec a}|^2$ contributions
of (\ref{trBlnDfinal}), and subtracting the eight order $|\dot{\vec a}|^2$
terms in (\ref{trBlnDfinal}) with $\cos\beta a_0\to-\cos\beta a_0$, we arrive
finally at the effective action
\bea
&&S_{\rm eff}=\int\limits_0^\beta d\tau~\left\{\frac{T_0}4\,v^2
+\frac8\beta\,\ln\frac{\cosh\frac{\beta r}{2\pi\alpha'}-\cos\beta a_0}
{\cosh\frac{\beta r}{2\pi\alpha'}+\cos\beta a_0}\right.\non&&+\,
\frac{v^2}{12r}\,\frac1{\cosh^2\frac{\beta r}{2\pi\alpha'}-\cos^2\beta a_0}
\non&&\times\,\left[38\,\frac{2\pi\alpha'}{r^2}\,\sinh\frac{\beta r}
{2\pi\alpha'}\,\cos\beta a_0\right.\non&&-\,
\frac{38\beta\cosh\frac{\beta r}{2\pi\alpha'}
\,\cos\beta a_0}r\,\frac{2-\cosh^2\frac{\beta r}{2\pi\alpha'}-\cos^2\beta a_0}
{\cosh^2\frac{\beta r}{2\pi\alpha'}-\cos^2\beta a_0}\non&&+\,
\frac{2\beta^2}{2\pi\alpha'}\,\frac{\sinh\frac{\beta r}{2\pi\alpha'}\,
\cos\beta a_0}{\left(\cosh^2\frac{\beta r}{2\pi\alpha'}-\cos^2\beta a_0
\right)^2}\left(\cosh^4\frac{\beta r}{2\pi\alpha'}\right.\non&&-\,
\cos^2\beta a_0\left(2-\cos^2\beta a_0\right)\non&&-\left.\left.\left.
6\cosh^2\frac{\beta r}{2\pi\alpha'}
\,\sin^2\beta a_0\right)\right]+\dots\right\} \ ,
\label{Sefffinal}\eea
where the D0-brane separation $r$ is time-dependent and obeys periodic boundary
conditions on the temperature circle. The quantity $\frac{T_0}2$ is the reduced
mass of the two D-particle system, while $\frac r{2\pi\alpha'}$ is the energy
of a string which has Dirichlet boundary conditions on hypersurfaces a distance
$r$ apart. Note that the effective action is an odd function of the variable
$x=\cos\beta a_0$. The action (\ref{Sefffinal}) simplifies in the limit $\beta
r\gg2\pi\alpha'$, and to leading orders in the low-temperature expansion we
have
\bea
&&S_{\rm eff}=\int\limits_0^\beta d\tau~\left[\frac{T_0}4\,v^2
-\frac{16}\beta~\e^{-\beta r/2\pi\alpha'}\,\cos\beta a_0\right.\non&&+\,
\frac{v^2}{6r}\,\cos\beta a_0~\e^{-\beta r/2\pi\alpha'}\left(19\,
\frac{2\pi\alpha'}{r^2}+\frac{19\beta}r+\frac{\beta^2}{2\pi\alpha'}\right)
\non&&+\biggl.{\cal O}\left(\e^{-3\beta r/2\pi\alpha'}\,\cos^3\beta a_0\right)+
\dots\biggr] \ .
\label{SefflowT}\eea

The second term in (\ref{Sefffinal}) has a direct interpretation in string
perturbation theory. One can compute the annulus diagram for the open
superstring, in compactified Euclidean time of circumference $\beta$, whose
ends lie on two stationary D0-branes separated by distance $r$. The charges at
the endpoints of the string couple to a constant $U(1)$ gauge field which is
parametrized by $\nu\in(-1,1]$, and which enters the problem through the
quantized temporal momentum $p^0=2\pi(n-\nu)/\beta$, $n\in\zed+\frac{a-1}2$, of
the open string whose worldsheet winds around the spacetime cylinder. Then, the
one-loop thermal partition function of the string gas can be written as
\cite{GreenT}
\be
Z_{\rm str}(\beta,r,\nu)=\prod_{N=0}^\infty\left|\frac{1+\e^{-\beta E_N+i
\pi\nu}}{1-\e^{-\beta E_N+i\pi\nu}}\right|^{2d_N} \ ,
\label{Zstr}\ee
where the superstring spectrum is given by
\be
\sqrt{\alpha'}E_N=\sqrt{\frac{r^2}{4\pi^2\alpha'}+N} \ ,
\label{stringspec}\ee
with $N$ the oscillator occupation number, and $d_N$ is the degeneracy of
superstring states at level $N$ which may be computed from the generating
function
\be
8\prod_{n=1}^\infty\left(\frac{1+\e^{-nl}}{1-\e^{-nl}}\right)^8
=\sum_{N=0}^\infty d_N~\e^{-Nl} \ .
\label{dN}\ee
For the lowest level we have $d_0=8$ and $E_0=r/2\pi\alpha'$. The factor of 2
in the power of (\ref{Zstr}) is again due to the exchange symmetry of the
string endpoints. The partition function (\ref{Zstr}) is equal to the ratio of
the Fermi and Bose distributions with power (twice) the degeneracy of states
and the parameter $i\nu$ playing the role of a chemical potential. The static
limit $v=0$ of (\ref{Sefffinal}) coincides with $\ln Z_{\rm str}$ truncated to
the massless modes ($N=0$) with the identification $\pi\nu=\beta a_0$.

As stressed in \cite{AMS}, the integration over $a_0$ of the effective action
is required for gauge invariance of the free energy, or equivalently to enforce
Gauss' law for the charges at the ends of the open string which are induced on
D-branes. The effective potential (\ref{SeffD0}) between D0-branes is thereby
given from (\ref{Sefffinal}) as $S_{\rm eff}[\vec
y^{\,a}]=-\ln\int_{-1}^1d\nu\,\e^{-S_{\rm eff}}$. The reason has a natural
explanation in the closed string formulation, obtained by mapping the open
string annulus diagram onto the cylinder diagram via the standard modular
transformation. Then, the path integral describes the closed string propagator
corresponding to the interaction between two D0-branes, rather than the thermal
partition function as in the case of an open string. When two D0-branes
interact, they can exchange several closed strings, not only one. As all such
exchanges are of the same order in the string coupling constant, they
exponentiate since the closed strings are identical and naturally produce the
result (\ref{Zstr}) in the closed string language. Furthermore, in this
formulation it is clear that there is only a single gauge field parameter $\nu$
for each multi-string term, because now the system is composed of just two
interacting D0-branes rather than a gas of D-particles. These facts result in
the effective potential as claimed.

In the static limit, this
potential is logarithmic and attractive at short distances. The singularity
occurs as the D0-branes fall on top of one another, in which case the
non-abelian gauge symmetry which is broken by separated branes is restored
\cite{Witten1}. Then, the one-loop approximation breaks down, and this
demonstrates that the thermodynamics of D0-branes must be treated as a problem
in {\it quantum} statistical mechanics, defined by the path integral
(\ref{ZYM}) over both periodic and anti-periodic trajectories $\vec y(\tau)$.
On the other hand, the leading velocity corrections to the static thermal
potential are repulsive at short distances. Note that these corrections are of
order $v^2$ and vanish in the zero temperature limit, as expected. This
illustrates that the moduli space of the two D-particle system is curved in a
very non-trivial way by thermal effects. The calculation presented in this
subsection can be extended to compute the thermal corrections to the order
$v^4$ gravitational interaction between moving D-branes, and thereby to shed
more light on the dynamical role of D0-branes in black hole thermodynamics.

\section{Noncommutative D-brane geometries}

A recent surge of activity in string theory and quantum field theory has come
with the realization that D-branes in certain background supergravity fields
lend an explicit realization to some old ideas that the classical notions of
spacetime and general relativity at very short distances must be drastically
altered. At these length scales quantum gravitational fluctuations cannot be
ignored and spacetime is no longer described by a differentiable manifold.
Following earlier suggestions, noncommutative geometry has been proposed as the
appropriate mathematical framework to describe the short scale structure of
spacetime, and in particular non-perturbative properties of string theory. The
fact that quantum field theory on a noncommutative space arises naturally in
string theory \cite{SW1} and M-theory \cite{CDS} suggests that spacetime
noncommutativity is a general feature of a unified theory of quantum gravity.

D-brane worldvolumes become noncommutative manifolds when there is a constant
Neveu-Schwarz two-form field ${\cal B}_{\mu\nu}$ on them. This field can be
coupled to the usual open string $\sigma$-model (\ref{Sstrip}) in the neutral
limit $e_1=e_2=e$ by adding the topological action $-i\int_\Sigma x^*{\cal B}$.
This term is a total derivative and so it only contributes to the boundary
conditions on the string fields, not to their equations of motion. The
endpoints of the string (the boundaries $\sigma=0,1$ of $\Sigma$) are now
interpreted as ending on a D-brane of a certain dimensionality. The $\cal B$
field only appears in a gauge invariant combination with the $U(1)$ gauge field
on the brane as $B_{\mu\nu}={\cal B}_{\mu\nu}-eF_{\mu\nu}$. Therefore, the
uniform Neveu-Schwarz background field is equivalent to a constant
electromagnetic field on the D-brane and so the following analysis will unify
our discussions from earlier sections. Since the background fields can be
gauged away in the directions transverse to the D-brane worldvolume, we shall
only study the quantities associated with the worldvolume hyperplane itself.

When the target space has Euclidean signature, the noncommutativity of the
string endpoint coordinates can be understood through the analogy, discussed in
section II.B.2, between the external field problem for open strings and the
classic Landau problem. The $\sigma$-model action describing the coupling of
strings to a magnetic field on the branes is formally equivalent to the Landau
action
\be
S_{\rm L}=\int d\tau~\left(\frac M2\,\dot{\vec y}^{\,2}-Q\,
\dot{\vec y}\cdot\vec A\,\right) \ ,
\label{Landau}\ee
which describes the motion of a particle of charge $Q$ and mass $M$ in the
plane $\vec y=(y^1,y^2)$ and in the presence of a uniform perpendicular
magnetic field of magnitude $B$. Here $A_i=-\frac B2\,\epsilon_{ij}\,y^j$ is
the corresponding vector potential. In the limit of a strong magnetic field
$B\gg M$ (with $M$ fixed), the action (\ref{Landau}) is already expressed in
phase space with the spatial coordinates $y^1,y^2$ being the canonically
conjugate variables. In canonical quantization, the position variables become
noncommuting operators with $[y^i,y^j]=(i/QB)\,\epsilon^{ij}$. The mass gap
between Landau levels is $QB/M$, so that the limit of strong magnetic field
projects the quantum mechanical spectrum of this system onto the lowest Landau
level and the spatial coordinates live in a noncommutative space. As we will
see in the following, this is precisely what happens to the string endpoints
when there is a constant magnetic field on the D-branes, and the D-brane
worldvolume becomes a noncommutative manifold. However, as one can anticipate
from our earlier analyses, the picture changes drastically in Minkowski
signature corresponding to an electric field on the branes.

\subsection{Magnetic fields and noncommutative field theory}

We will start with the case of Euclidean spacetime, so that $B_{\mu\nu}$
represents a uniform magnetic field on the D-branes, which we assume is of
maximal rank. The open string boundary conditions are given by (\ref{NDbcs})
with the replacements of $-e_kF_{\mu\nu}$ by $B_{\mu\nu}$ everywhere. To see
how noncommutative geometry arises on the D-brane worldvolume, we will use the
operatorial, covariant quantization formalism of section II.B.1, but now in
full generality and with a more careful analysis of the canonical quantization.
The mode expansions which solve the bulk equations of motion $\Box\,x^\mu=0$
and the boundary conditions (\ref{NDbcs}) are given by the familiar expressions
\bea
x^\mu(\tau,\sigma)&=&y^\mu\non&&+\,
2\alpha'\left(G^{-1}\right)^{\mu\nu}\left(q_\nu\,
\tau-2\pi^2\alpha'B_\nu^{~\lambda}\,q_\lambda\,\sigma\right)\non&&
+\,\sqrt{2\alpha'}\,\sum_{n\neq0}\frac{\e^{-in\tau}}n\,\Bigl(i\,a_n^\mu\cos
n\pi\sigma\Bigr.\non&&-\Bigl.2\pi\alpha'B^\mu_{~\nu}\,
a_n^\nu\sin n\pi\sigma\Bigr) \ ,
\label{NCmodes}\eea
where
\be
G=\id-(2\pi\alpha'B)^2
\label{NCBIfactor}\ee
is the usual Born-Infeld factor. As is evident from the expression for the
string propagator in the background field \cite{SW1}, the symmetric tensor
(\ref{NCBIfactor}) is the {\it open} string metric, i.e. the metric seen by the
endpoints of the string, while $\id$ is the bulk, closed string metric that
defines the $\sigma$-model action.

We can now straightforwardly compute the equal-time, canonical commutation
relations as described in section II.B.1. Those involving the worldsheet
momentum density uniquely fix the usual Heisenberg commutation relations for
the zero modes $x^\mu,q_\nu$ and the standard Heisenberg-Weyl commutation
relations for the oscillatory modes $a_n^\mu$ in the metric $G$. The subtle
relation comes from the equal-time commutator
$[x^\mu(\tau,\sigma),x^\nu(\tau,\sigma')]=0$ \cite{Ardalan,ChuHo}. By using the
Heisenberg-Weyl commutation relations and the mode expansion (\ref{NCmodes}),
this commutator is readily seen to be given by
\bea
\Bigl[x^\mu(\tau,\sigma)\,&,&\,x^\nu(\tau,\sigma')\Bigr]=
\left[y^\mu,y^\nu\right]-i\,\theta^{\mu\nu}\left(\sigma+\sigma'\right)
\non&&~~-\,i\,\theta^{\mu\nu}\,\sum_{n\neq0}\frac1n\,\sin n\pi
\left(\sigma+\sigma'\right) \ ,
\label{0commrel}\eea
where
\be
\theta=-(2\pi\alpha')^2\,G^{-1}B
\label{thetanoncomm}\ee
is the open string external field. By integrating the completeness relation
(\ref{coscompl}) we may arrive at the Fourier series expansion
\be
\sum_{n\neq0}\frac1n\,\sin n\pi\left(\sigma+\sigma'\right)=1-\left(
\sigma+\sigma'\right)
\label{sincompl}\ee
for $\sigma+\sigma'\in(0,2)$. From (\ref{0commrel}) we see that for
$\sigma,\sigma'\in(0,1)$ in the {\it bulk} of the string worldsheet, the
canonical commutation relations may be satisfied by fixing the commutators of
the zero mode position operators to be
\be
[y^\mu,y^\nu]=i\,\theta^{\mu\nu} \ .
\label{noncommrel}\ee
The $y^\mu$ therefore generate a noncommutative algebra of operators and are
interpreted as coordinates on a noncommutative space. They guarantee that the
equal-time commutators are unmodified in the bulk of the worldsheet. This must
be the case, since the coupling to the external field only modifies the
boundaries of the string worldsheet, not the interior.

However, from (\ref{0commrel}) and (\ref{noncommrel}) it now follows that the
open string endpoint coordinates become noncommuting operators,
\bea
\Bigl[x^\mu(\tau,0)\,,\,x^\nu(\tau,0)\Bigr]&=&i\,\theta^{\mu\nu} \ , \non
\Bigl[x^\mu(\tau,1)\,,\,x^\nu(\tau,1)\Bigr]&=&-i\,\theta^{\mu\nu} \ ,
\label{endptnoncomm}\eea
with all other embedding field commutators vanishing. The commutation relations
(\ref{endptnoncomm}) arise from the compatibility of the open string boundary
conditions with the standard commutators, and they imply that the presence of
the $B$-field deforms the D-brane worldvolume to a noncommutative manifold.
Note that the noncommutativity of the worldvolume coordinates cannot be probed
by any closed string objects (such as supergravity fields). This is because of
the flip in sign between the commutators (\ref{endptnoncomm}) at the two ends
of the string which arise from the change of orientation. The left and right
moving modes receive equal and opposite contributions from the $B$ field and
the noncommutativity averages out in the region between the two D-branes. In
fact, one can explicitly calculate that the open string center of mass
coordinates $x_{\rm cm}^\mu(\tau)=\int_0^1d\sigma\,x^\mu(\tau,\sigma)$ commute.
Thus the transverse space remains an ordinary (commutative) manifold. Indeed,
we recall that the external field in the neutral case does not change the
physical spectrum of the theory. The only effect of the magnetic field is the
change the metric $\id$ to the open string metric (\ref{NCBIfactor}). Recall
also from section II.C that the open string is not point-like, but rather
behaves like a neutral magnetic dipole whose two endpoints are at different
positions. The dipole grows in the direction transverse to the motion by an
amount proportional to $B_{\mu\nu}\,q^\nu$, and the fuzziness of space
originates from its size \cite{Sh-J,BigSuss}.

It is also possible to see noncommutativity in the charged string case
\cite{Chu}, which formally corresponds to different external fields $B_k={\cal
B}-e_kF$ on the two D-branes between which the open strings stretch. By using
the mode expansions (\ref{xpmchargedmodes}) and the canonical commutation
relations (\ref{anpmcomm},\ref{ypmcomm}) we find
\bea
&&\Bigl[x^+(\tau,\sigma)\,,\,x^-(\tau,\sigma')\Bigr]=\frac1{2\alpha'(B_1-B_2)}
\non&&~~+\,\sum_{n=-\infty}^\infty\frac1{n-\alpha}\,\cos\Bigl((n-\alpha)\pi
\sigma+\arctan2\pi\alpha'B_1\Bigr)\non&&~~\times\,\cos\Bigl((n-\alpha)\pi
\sigma'+\arctan2\pi\alpha'B_1\Bigr) \ .
\label{xpmchargednoncomm}\eea
By using the identity
\be
\sum_{n=1}^\infty\frac{2\alpha}{\alpha^2-n^2}+\frac1\alpha=\pi\cot\pi\alpha
\label{cotid}\ee
for $\alpha\notin\zed$, we may infer the noncommutativity relations
\bea
\Bigl[x^+(\tau,0)\,,\,x^-(\tau,0)\Bigr]&=&i\,\theta_1\non
\Bigl[x^+(\tau,1)\,,\,x^-(\tau,1)\Bigr]&=&-i\,\theta_2
\label{xpmnctheta}\eea
with all other embedding field commutators vanishing \cite{Chu}, where
$\theta_k=(2\pi\alpha')^2B_k[1-(2\pi\alpha'B_k)^2]^{-1}$. Thus the
noncommutativity is localized at the string endpoints and is determined by the
field strengths on the D-branes. Exactly the same noncommutativity factors are
obtained as if one quantized an individual open string ending on the same
D-brane. This simply reflects the fact that noncommutativity is an intrinsic
property of the brane worldvolume and not of the short-distance probe that is
used. Notice also that the
noncommutativity parameters are proportional to the string
scale $\alpha'$ and thereby represent genuine stringy effects. The results are
in fact exact to all orders in $\alpha'$ and the string coupling constant
$g_s$, because noncommutativity is a short distance effect which doesn't care
about the worldsheet topology. The loop corrections to the above results have
been analysed in \cite{Laidlaw} with the same conclusions.

These same results can be reached by studying operator product expansions
of open string vertex operators \cite{Schomerus,SW1}. In this analysis one can
identify a particular regime of the string theory in which the vertex operator
algebra reduces to a deformation of the ring of functions $f(y)$ on the D-brane
worldvolume \cite{LLS,SW1}. It corresponds to taking the correlated limits
$\alpha'\to0$ (the field theory limit), $g_s\to0$ (weakly coupled strings), and
$B_{\mu\nu}\to\infty$ (strong magnetic field), with the quantities
$(\alpha')^2B_{\mu\nu}$ and $g_s\,\sqrt{\det\alpha'B}$ finite. The open string
metric (\ref{NCBIfactor}) is given by $G=-(2\pi\alpha'B)^2$ in this limit,
since the closed string metric effectively scales out of the problem.
Furthermore, from (\ref{NCmodes}) it follows that the massive string modes are
also scaled away from the endpoint zero modes. Thus all closed string states
are completely decoupled from the problem, i.e. the gravitational modes are
removed and an effective field theory remains. However, this is not a
conventional field theory, because the noncommutativity parameter is also
finite in this limit, $\theta=1/B$. Indeed, because of (\ref{noncommrel}), the
resulting projection of the vertex operator algebra is not an ordinary function
algebra, but rather that which is obtained by deforming the pointwise
multiplication $f(y)g(y)$ of two functions to a product defined by a
bi-differential operator of infinite order \cite{LLS,SW1}. This is is given by
the classic Moyal star-product
\be
f(y)\star g(y)=\left.\exp\left(\frac i2\,\theta^{\mu\nu}\,\frac\partial
{\partial y^\mu}\,\frac\partial{\partial y'^\nu}\right)f(y)g(y')
\right|_{y'=y}
\label{Moyal}\ee
which is associative, but non-local and noncommutative. The commutation
relations (\ref{noncommrel}) may then be satisfied by replacing ordinary
operator products with star-products of the coordinates $y^\mu$. Note that in
this decoupling limit the $\sigma$-model action (\ref{Sstrip}) reduces to a sum
of two quantum mechanical, boundary actions for the endpoint charges which are
each formally equivalent to the Landau action (\ref{Landau}) in the limit
$B\to\infty$. This limit is thereby analogous to the projection onto the
lowest-lying Landau level. The effects of noncommutativity from the string zero
modes is emphasized in \cite{Kar2}.

Proceeding as before, it can be shown that the effective field theory is given
by a noncommutative generalization, obtained by replacing ordinary
(commutative) products of fields with the Moyal product (\ref{Moyal}), of the
Dirac-Born-Infeld action \cite{Leigh} on the D-brane worldvolume which
describes non-linear electrodynamics on a fluctuating membrane \cite{Lee}. This
can be used to identify the effective open string coupling constant as
\cite{SW1}
\be
G_s=g_s\,\sqrt{\det(\id+2\pi\alpha'B)} \ .
\label{Gs}\ee
After supersymmetrization, the low-energy effective field theory of
noncommutative D-branes is noncommutative supersymmetric Yang-Mills theory with
16 supercharges (the number of supersymmetries preserved by the D-branes in the
$B$ field background) and spacetime metric $G_{\mu\nu}$. The Yang-Mills
coupling constant in the decoupling limit described above is given by $g_{\rm
YM}^2\propto G_s=g_s\,\sqrt{\det2\pi\alpha'B}$.

Quantum field theory on a noncommutative space appears to be the unique
consistent deformation of ordinary quantum field theory. These theories exhibit
a variety of novel effects which lead to new physics that are not encountered
in conventional quantum field theories. Many of these effects have counterparts
in string theory, and noncommutative field theories are believed to lie
somewhere between ordinary field theory and string theory. For instance, one of
the most important results is that infrared and ultraviolet effects do not
decouple in a noncommutative field theory \cite{MvRS}, which can be understood
from the fact that the open string dipoles grow in size with their energy. The
larger the momentum, the larger is the spatial extension of the object.
Furthermore, noncommutative scalar field theories can contain stable soliton
solutions even if their commutative counterparts don't \cite{GMS}, and these
noncommutative solitons can be realized as D-branes in string field theory.
Because of these striking features, intensive studies have been initiated which
use noncommutative quantum field theory to study D-branes in the presence of a
background magnetic field.

\subsection{Electric fields and noncommutative open string theory}

Let us now Wick rotate to Minkowski signature and consider a uniform electric
field $E=|\vec E\,|$ on the branes, i.e. $B_{ij}=0$ and $E_i=-iB_{0i}\neq0$.
Then $\theta^{0i}\neq0$ and the D-brane worldvolume is space/time
noncommutative. There are several reasons why one is interested in such a
noncommutative theory. First of all, the lack of commutativity of time is in
conflict with our current understanding of quantum mechanics, where time is not
an operator but rather a parameter which labels the evolution of the system.
Understanding space/time noncommutativity may therefore shed light on the role
of time in string theory and quantum gravity. Secondly, the space/time
commutator implies the uncertainty relation $\Delta y^0\,\Delta
y^i\sim\theta^{0i}$ between time and space. This is simply the string
space/time uncertainty principle that has been advocated as a generic property
of string theory \cite{Yoneya}. Finally, in the absence of external fields the
effective supersymmetric Yang-Mills theory on the four-dimensional worldvolume
of coincident D3-branes is known to possess an exact Montonen-Olive $S$-duality
$g_{\rm YM}\leftrightarrow1/g_{\rm YM}$. In the presence of a background
electromagnetic field we expect this symmetry to act as an electric-magnetic
duality exchanging electric and magnetic degrees of freedom. Naively then, we
expect that the strong coupling dual of spatially noncommutative Yang-Mills
theory in four dimensions to be a temporally noncommutative gauge theory.

This latter line of reasoning is, however, incorrect. Noncommutative quantum
field theory with a noncommuting time direction is neither unitary nor causal.
It suffers from severe acausal effects such as events which precede their
causes and objects which grow instead of Lorentz contract as they are boosted.
For example, the open string electric dipoles extend longitudinally by an
amount proportional to $\vec E\cdot\vec q$. However, the string theory in a
background electric field is, at least perturbatively, unitary and causal, as
is evident in first quantization. Stringy effects eventually conspire to cancel
the acausal effects that arise (for instance in the zero mode dipoles), and the
model at the level of string theory is perfectly well-defined. Therefore, while
the $S$-dual of the electric field problem for open strings is certainly the
corresponding magnetic one, the noncommutative Yang-Mills theories cannot be
related in such a manner. What has gone wrong is that the electric field
problem does not possess a noncommutative field theory limit \cite{SLN,GMMS}.
Recall from the previous subsection that one of the decoupling limits involved
making the external magnetic field arbitrarily large. In the electric case, the
system destabilizes above the critical value $E_c$. This instability is now
reflected in the singularities that arise at $E=E_c$ in the open string
parameters (\ref{NCBIfactor}), (\ref{thetanoncomm}) and (\ref{Gs}). It prevents
the correlated limit of the previous subsection from being taken. The electric
field cannot be scaled to infinity, and so one cannot reach the field theoretic
limit $\alpha'\to0$ in which all string oscillator modes decouple.

The key point though is that the effective tension of an open string stretched
along the direction of the electric field is given by (\ref{Teff}). One can
take a limit in which $\alpha'\to0$, and the theory is space/time
noncommutative and decouples the bulk modes, including gravity, off the branes.
However, the effective string scale $\alpha'_{\rm eff}=1/2\pi T_{\rm eff}$ is
finite in this limit and the effective theory will be a string theory, not a
field theory. For this, we rotate so that the electric field lies along the
1-axis, and rescale the coordinates so that the diagonal elements of the closed
string metric in the 0--1 plane are proportional to
$[1-(2\pi\alpha'E)^2]^{-1}$. We then take the limit whereby the electric field
becomes critical, $2\pi\alpha'E\to1$, and $\alpha'\to0$ with $\alpha'_{\rm
eff}$ fixed. For finite $\alpha'$, the open strings are effectively tensionless
in the limit $E\to E_c$ and the open string metric (\ref{NCBIfactor}) is
finite. The closed string metric scales to infinity, and the Moyal phases are
determined by the effective string scale as $\theta=2\pi\alpha'_{\rm eff}$ and
are therefore finite. Recall now that the neutral open string spectrum is
unaltered by the electric field, and so the open string states are of finite
mass. Thus we are left with an open string theory on the D-brane worldvolume
which is a space/time noncommutative manifold. The fact that the
noncommutativity scale is intrinsically tied to the string scale means that in
order to make sense of a noncommutative space/time manifold, one needs to make
precise the notion of an Einstein spacetime at the string scale.

The main property of this string theory is that, unlike ordinary string theory
which requires closed string states for its consistency, it is completely
decoupled from the bulk worldsheet states. To see this, suppose that a light
open string state tries to escape to the bulk by turning into a closed string
state (via a modular transformation). For this to occur, the stretched open
string has to bend over in order for its endpoints to touch each other. Part of
it will stretch against the electric field and will thereby become very heavy
as $E\to E_c$. Thus the closed string modes become infinitely massive, and
energetics prevent the open strings which live on the branes from turning into
closed strings and propagating into the bulk. Note that, according to
(\ref{Gs}), this string theory is interacting provided we scale the closed
string coupling $g_s\to\infty$. Therefore, these open strings describe a
particular limit of strongly coupled closed strings in a critical electric
field.

We conclude that in the low energy limit considered above, the effective theory
includes interacting open strings on the D-branes together with decoupled free
closed strings in the bulk region. The open string theory is decoupled from
gravity, and the underlying spacetime on the D-branes is noncommutative. This
theory is known as noncommutative open string theory \cite{SLN}. In the case of
D3-branes it is the strong coupling dual of supersymmetric noncommutative
Yang-Mills theory in four dimensions \cite{GMMS}. The action involving these
open strings is related to the action of ordinary open string theory by the
replacement of all ordinary products of string fields with the appropriate
noncommutative Moyal products (\ref{Moyal}). The thermal ensembles, and in
particular the Hagedorn behaviour \cite{GGKRW}, of this string theory are
particularly interesting since this theory does not contain closed strings and
decouples from gravity. In the conventional superstring theory, which is
difficult to study because of the thermodynamic instabilities that arise in
gravitating systems, there is a first order phase transition below the Hagedorn
temperature \cite{AW}. In the present case, one finds that, in the scaling
limit and as the temperature is increased, a massless closed string state
appears in the bulk at precisely the Hagedorn temperature (\ref{THE}) arising
from the open string density of states. The Hagedorn transition in this case is
a second order phase transition, and the high temperature phase involves long
fundamental strings separating from the D-branes on which the noncommutative
open string theory is defined \cite{GGKRW}.

\acknowledgments

The work of J.A., Y.M.M. and G.W.S. was supported in part by ``MaPhySto'',
{\it Center for Mathematical Physics and Statistics}, financed by the Danish
National Research Foundation. J.A. and R.J.S. acknowledge support from the EU
network ``Discrete Random Geometry'', grant HPRN--CT--1999--00161, and from
the ESF network no. 82 on ``Geometry and Disorder''. The work of R.J.S. was
supported in part by an Advanced Fellowship from the Particle Physics and
Astronomy Research Council (U.K.).

\end{document}